\documentclass[10pt,prd,aps,nofootinbib,eqsecnum,notitlepage,nobibnotes,superscriptaddress,preprintnumbers,twocolumn]{revtex4-1}
\usepackage{hyperref}

\usepackage[a4paper, left=20mm,right=20mm,top=20mm,bottom=20mm]{geometry}
\usepackage[titletoc,toc,title]{appendix}
\usepackage{soul}
\usepackage{xcolor}

\usepackage{mathtools}
\usepackage{tensor}
\usepackage{float}
\usepackage{derivative}
\usepackage{bm}
\usepackage{tabularray}
\usepackage{suffix}

\usepackage{amsmath, amsfonts,bm}
\usepackage{graphicx}
\usepackage{xcolor}
\usepackage[normalem]{ulem}

\usepackage[dvipsnames]{xcolor}
\usepackage{physics}
\usepackage{placeins}
\usepackage{cancel}

\DeclareMathAlphabet{\mathup}{OT1}{\familydefault}{m}{n}

\def\dd{\mathrm{d}}

\renewcommand\vec[1]{\bm{#1}}

\newcommand{\be}{\begin{equation}} 
\newcommand{\ee}{\end{equation}}

\begin{document}
\title{Circular stable orbits in $f(R)$ realistic static and spherically-symmetric spacetimes
}
\author{N\'estor Rivero Gonz\'alez}
    \email{nestoriv@ucm.es}
    \affiliation{Departamento de F\'{\i}sica Te\'orica, Ciudad Universitaria, Universidad Complutense de Madrid, 28040 Madrid, Spain.}
    \author{\'{A}lvaro de la Cruz-Dombriz}
      \affiliation{ Departamento de F\'{i}sica Fundamental, Universidad de Salamanca, 37008 Salamanca, Spain}
    \affiliation{Cosmology and Gravity Group, Department of Mathematics and Applied Mathematics, University of Cape Town, Rondebosch 7700, Cape Town, South Africa}
    \author{Gonzalo J. Olmo}
\affiliation{Instituto de F\'isica Corpuscular (IFIC), CSIC‐Universitat de Val\`{e}ncia, Spain}
\affiliation{Universidade Federal do Cear\'a (UFC), Departamento de F\'isica,\\ Campus do Pici, Fortaleza - CE, C.P. 6030, 60455-760 - Brazil.}

\date{\today}
\pacs{04.50.Kd, 98.80.-k, 98.80.Cq, 12.60.-i}


\begin{abstract}
We investigate the geodesic structure of realistic static and spherically symmetric spacetimes embedding neutron stars in metric $f(R)$ gravity, focusing on the quadratic Starobinsky model $f(R)=aR^2$ with $a<0$. Neutron-star solutions are obtained by numerically solving the modified Tolman--Oppenheimer--Volkoff system for several realistic equations of state. Such solutions  are then matched consistently to the exterior vacuum geometry by enforcing the full set of junction conditions required in metric $f(R)$ theories. Using an effective potential approach, we show that stable circular orbits appear in discrete radial bands separated by forbidden regions, with a dominant principal band of stability that depends sensitively on the stellar central pressure, the equation of state, and the magnitude of the parameter $|a|$. Outside the stable bands, massive particles can have bound but unstable precessing trajectories as well as unbounded motions. On the other hand, for null geodesics, we find no evidence for photon spheres outside the neutron star within the parameter range studied.
\end{abstract}


\maketitle

\section{Introduction}
\label{sec:intro}

Throughout the last century, the motion of test particles and light rays in strong gravitational fields has provided some of the most powerful probes of the underlying gravitational theory. For instance, in the context of the theory of General Relativity (GR), the geodesic structure of static and spherically symmetric spacetimes is well understood and has played a central role in the description of compact objects, accretion processes, and observational signatures such as gra\-vi\-ta\-tional lensing and quasi-periodic oscillations 
\cite{Chandrasekhar:1985kt, Shapiro:1983du}. On the other hand, in theories beyond GR, modifications of the gravitational dynamics can significantly alter the spacetime geometry and, consequently, the behavior of both null and timelike geodesics. 
The study of particle trajectories therefore offers a va\-lua\-ble window into the phenomenology of modified, also often dubbed extended, theories of gravity and provides a deeper understanding of their physical implications \cite{Clifton:2011jh,Nojiri:2010wj,
Capozziello:2011et}.

Motivated by the role that particle trajectories play in probing gravitational physics, considerable attention has been devoted in recent years to the analysis of geodesic motion in alternative gravitational frameworks, including metric and Palatini formulations of $f(R)$ gravity, Ricci-based theories, massive gravity, and extensions involving nonlinear matter sources, e.g., \cite{Olmo:2011uz,BeltranJimenez:2017doy, Olmo:2020fri,Afonso:2017bxr,Olmo:2015wda, Olmo:2015nhk,Afonso:2018bpv} and references therein.
In particular, numerous works have shown that extended gravity theories can lead to profound modifications of the causal and geodesic structure of spacetime. These include the replacement of curvature singularities by wormhole-like geometries, as in \cite{Olmo:2015wda}, the emergence of geodesically complete black-hole solutions \cite{RubieraGarcia:2015oea}, and the appearance of effective potential barriers that strongly affect the propagation of ma\-ssive particles and photons \cite{Olmo:2015wda, Olmo:2015nhk,Afonso:2018bpv}. Both null and timelike geodesics have been systematically analyzed in these and related settings, revealing novel features absent in GR, such as repulsive cores, disconnected regions of motion, and modified sta\-bi\-lity properties of circular orbits \cite{Guerrero:2020crm, DeLaurentis:2018udl}. 
In this regard, within the framework of 
$f(R)$ gravity and as natural extensions of the seminal works by Tolman-Oppenheimer-Volkoff \cite{oppenheimer_volkoff_1939, tolman_1939}, both the correct procedure for solving the re\-le\-vant field equations and the consistent embedding of realistic stellar models into static, spherically symme\-tric exterior spacetimes have been extensively studied; see, e.g., \cite{Olmo:2019flu,AparicioResco:2016xcm, Astashenok:2017dpo,
arapoglu_deliduman_eksi_2011, astashenok_capozziello_odintsov_2013,
yazadjiev_doneva_kokkotas_staykov_2014,
capozziello_delaurentis_farinelli_odintsov_2016,
Nashed:2021sji, Fernandez:2025cfp}.

Beyond the aforementioned frameworks, other mo\-di\-fied gra\-vi\-ty theories such as Einsteinian cu\-bic gra\-vi\-ty and massive gra\-vi\-ty also predict distinctive modifications of geodesic motion around compact objects. For instance, in Einsteinian cubic gravity, higher-curvature couplings modify the location and properties of bound orbits and innermost stable circular orbits \cite{Li:2024tld}. Similarly, massive gravity theories yield black-hole solutions with altered orbital dynamics due to the graviton mass, affecting the structure of stable and unstable orbits \cite{Fathi:2023bbh}. Also, other approaches, including effective quantum-corrected black holes, demonstrate how quantum gravity corrections can influence periodic orbits and gravitational waveform signatures in orbiting motion \cite{Chen:2025aqh,Bragado:2025jrg}. Photon propagation can also be severely affected by nonlinearities in the electromagnetic field equations, having deep implications for the causal structure of space-time \cite{dePaula:2024yzy} and the shadows of compact astrophysical objects \cite{daSilva:2023jxa}.

In addition to these studies, the dynamics of test particles around compact objects in modified gra\-vi\-ty has also been investigated outside strong-field black-hole scenarios, for instance in the presence of quintessence fields or under external magnetic influences, highlighting further rich phenomenology in galactic and cosmological contexts \cite{Wang:2023otn,Yang:2022rmp}. Effec\-tive potential methods and numerical integration of the geodesic equations have proven particularly useful in characterizing orbital dynamics and identifying stability regions in these scenarios \cite{Bragado:2025jrg}.

We thus see that a predominant feature in mo\-di\-fied theories of gravity beyond GR is the existence of new dynamical degrees of freedom, which can be encoded in scalar, vector, or tensor fields of different types. Such fields represent new forms of stress-energy and their distribution in spacetime is relevant to determine the dynamics of the corresponding systems. For localized matter distributions, such as stars capable of significantly affecting the curvature of spacetime, the presence of extra dynamical fields may be enhanced, affecting the orbital dynamics of nearby objects in ways that can help us detect their presence. For instance, a star surrounded by a matter scalar field may exhibit an effective mass that increases with the distance to its surface due to the energy stored in the scalar cloud. Since the stellar mass is expected to dominate over the scalar contribution, such effects are likely to be small. However, if the scalar degree of freedom is of gravitational origin, in addition to contributing to the total mass of the system, it can also have an impact on the effective Newton constant, which may decay in an exponential way or oscillate, depending on the sign of the effective squared mass of this degree of freedom. The observable impact of such oscillatory changes is likely to lead to clearer and more dramatic orbital features than exponential Yukawa-like decays, which require careful characterization. Compact stars, and neutron stars in particular, provide a natural arena to explore such effects, as their strong gravitational fields and well-understood microphysics make them sensitive probes of additional gravitational degrees of freedom.

Motivated by these considerations and recent
deve\-lop\-ments, in this work we investigate the motion of massive test particles in static and spherically sy\-mme\-tric spacetimes arising in metric $f(R)$ gravity, focusing on exterior solutions of realistic neutron star models. Unlike vacuum black-hole configurations, here we find that these spacetimes exhibit oscillatory behavior of metric functions outside the stellar surface, leading to nontrivial modifications of the effective potential governing particle motion 
\cite{Cooney:2009rr,Yazadjiev:2014cza,Capozziello:2015yza}. As we show, these features give rise to discrete radial bands of stable circular orbits, as well as novel orbital dynamics not present in GR. By analyzing the existence, stability, and structure of these orbits, and by comparing them with their GR counterparts, we aim to further clarify the phenomenology of particle motion in extended theories of gravity and to contribute to ongoing efforts to test gravity in the strong-field regime.

This paper is organized as follows. In Sec. \ref{sec:metricas} we introduce the theoretical framework of metric $f(R)$ gravity and specialize it in static and spherically symmetric spacetimes. We construct realistic neutron star configurations by solving the modified Tolman--Oppenheimer--Volkoff equations for several representative equations of state, and we impose the complete set of junction conditions required in metric $f(R)$ theories in order to obtain smoothly matched interior and exterior solutions without surface layers. Particular attention is paid to the behavior of the Ricci scalar and the resulting properties of the exterior vacuum geometry.
Then, in Sec. \ref{sec:massive orbits} we investigate the geodesic motion of massive test particles in the exterior spacetime.
First, in \ref{SecIIIA} we present the derivation of the conditions for the existence, boundedness, and stability of circular orbits using an effective potential approach, highlighting the emergence of discrete radial bands of stable motion. Subsequently, \ref{SecIIIB} and \ref{SecIIIC} analyze the properties of the principal stability band and the innermost stable circular orbit, examining their dependence on the stellar configuration and comparing them with their GR counterparts. The behavior of massive particles outside the stable regions is discussed in \ref{SecIIID}, while \ref{SecIIIE} introduces the geodesic equations and illustrates the resulting orbital dynamics through numerical integrations. Subsequently, in Sec. \ref{orbits_photons} we extend the previous analysis to null geodesics and examine the conditions for the existence of photon spheres in static and spherically symmetric $f(R)$ spacetimes.
To conclude, Sec.  \ref{Conclusions} contains a summary of our main results and discusses their implications for the geodesic structure of compact object spacetimes in metric $f(R)$ gravity. Throughout this work we use geometrized  units $c=G=1$ and adopt the metric signature  $\{+, -, -, -\}$.
\section{Analysis of static and spherically symmetric spacetimes in $f(R)$ theories}
\label{sec:metricas}

\begin{figure}[htbp!]
\begin{center}
    \centering
    \includegraphics[width=0.7\linewidth]{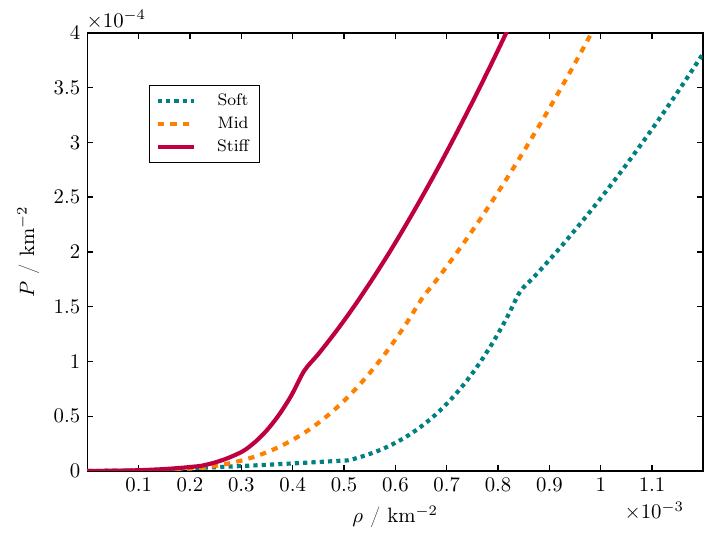}
    \caption{\footnotesize 
    ``Stiff'', ``middle'' and ``soft'' EOS based on potential models' data as in \cite{Hebeler:2013nza} that describe neutron matter. 
    Soft possesses an EOS in which pressure increases most slowly with density, and stiff is the one in which this growth is the most rapid. 
}
    \label{fig:EoS}
\end{center}
\end{figure}

To investigate astrophysically relevant solutions in $f(R)$ gravity, we focus on static and spherically symmetric configurations within the metric formalism, where the Levi-Civita connection is assumed and the action is thus varied with respect to the metric. The total action of the system is given by
\begin{equation}
    S = \frac{1}{2\kappa} \int {\rm d}^4x \sqrt{-g} \left[ R + f(R) \right] + S_M(g_{\mu\nu}, \phi),
\end{equation}
where $\kappa = 8\pi$ in natural units, $g$ is the determinant of the metric $g_{\mu\nu}$, $f(R)$ is a general function of the Ricci scalar, and $S_M$ denotes the matter action. Varying this action with respect to the metric yields the modified field equations:
\begin{equation}
\begin{aligned}
   R_{\mu\nu} - \frac{1}{2} &R g_{\mu\nu} = \frac{1}{1 + f_R} \bigg[ -\kappa T_{\mu\nu} - \nabla_\mu \nabla_\nu f_R \\
   &+g_{\mu\nu} \Box f_R + \frac{1}{2} (f(R) - R f_R) g_{\mu\nu} \bigg],    
   \label{fieldEQS}
\end{aligned}
\end{equation}
where $T_{\mu\nu}$ is the usual matter energy-momentum tensor, $f_R \equiv \frac{{\rm d}f}{{\rm }dR}$ and analogously for higher derivatives, and $\Box = \nabla^\alpha \nabla_\alpha$ is the d’Alembertian operator. In the static and spherically symmetric case, the usual four-dimensional line element takes the form:
\begin{equation}
\label{metric}
    {\rm d}s^2 = A(r)\, {\rm d}t^2 - B(r)\, {\rm d}r^2 - r^2\, ({\rm d}\theta^2 + \sin^2\theta\, {\rm d}\varphi^2),
\end{equation}
where $A(r)$ and $B(r)$ are metric functions to be determined by solving the field equations \eqref{fieldEQS}. For the matter sector, we consider a perfect fluid characterized by an energy-momentum tensor
\begin{equation}
    T_{\mu\nu} = (\rho + P)\, u_\mu u_\nu - P\, g_{\mu\nu},
\end{equation}
with $\rho$ the energy density, $P$ the pressure, and $u^\mu$ the four-velocity of the fluid. To close the system of equations, we shall consider in the following a barotropic equation of state (EoS)  of the form $\rho = \rho(P)$, reflecting the physical properties of the neutron star matter under consideration. In this work, three different EoS are considered, all extracted from \cite{Hebeler:2013nza}, and shown in Fig. \ref{fig:EoS}.

From the assumptions above, the field equations \eqref{fieldEQS} render the system \cite{AparicioResco:2016xcm}
\begin{equation}
\begin{aligned}
    A^{\prime\prime}= \;&\frac{A^{\prime}}2\left(\frac{B^{\prime}}B+\frac{A^{\prime}}A\right)+\frac{2B^{\prime}A}{rB}+\frac{2A}{(1+f_R)}\bigg[-\kappa BP\\
    &-\frac B2R+\left(\frac{A^{\prime}}{2A}+\frac2r\right)f_{2R}R^{\prime}-\frac B2f(R)\biggr],
\end{aligned}
\label{eq:de_A_f(R)}
\end{equation}
\begin{equation}
\begin{aligned}
    B^{\prime}= \;&\frac{2rB}{3(1+f_R)}\left[\kappa B(\rho+3P)+\frac{B}{2}R-\frac{3A^{\prime}}{2rA}+Bf(R)\right.\\
    &\left.-f_R\left(\frac B2R+\frac{3A^{\prime}}{2rA}\right)-\left(\frac3r+\frac{3A^{\prime}}{2A}\right)f_{2R}R^{\prime}\right],
\end{aligned}
\label{eq:de_B_f(R)}
\end{equation}
\begin{equation}
\begin{aligned}
    R^{\prime\prime} = \;&R^{\prime}\left(\frac{B^{\prime}}{2B}-\frac{A^{\prime}}{2A}-\frac{2}{r}\right)-\frac{B}{3f_{2R}}\Big[\kappa(\rho-3P)\\
    &-(1-f_R)R-2f(R)\Big]-\frac{f_{3R}}{f_{2R}} R^{\prime2},
\end{aligned}
\label{eq:de_R_f(R)}
\end{equation}
\begin{equation}
    P^{\prime}=-\frac{\rho+P}{2}\frac{A^{\prime}}{A},
\label{eq:de_P_f(R)}
\end{equation}
where the prime denotes derivative with respect to the radial coordinate. The last equation above becomes trivial on the star exterior since both pressure and energy density vanish there. To get physically acceptable solutions, one must integrate the system \eqref{eq:de_A_f(R)}-\eqref{eq:de_P_f(R)} in the radial coordinate together with the chosen EoS $\rho=\rho(P)$. Also, a  choice of initial conditions, such as 
\begin{equation}
     B(0)=1\,,\;     A'(0)=0\,,\; R'(0)=0\,,
\end{equation}
is required, as explained in \cite{AparicioResco:2016xcm}. The system closes with a finite central pressure $P(0)=P_c$. On the other hand,  as carefully explained in \cite{Senovilla:2013vra}, in the context of metric $f(R)$ gravity, the junction conditions for a compact star---excluding the existence of thin shells, double layers, etc.---impose the continuity of both $g_{\alpha\beta}$, $K_{\alpha\beta}$ (that is, the extrinsic curvature), $R$ and $\nabla_\alpha R$. Hence,\footnote{To study the conditions that must be satisfied at $\Sigma$, the fo\-llow\-ing notation is introduced:
$$
    [A] \equiv A(V^+)|_\Sigma - A(V^-)|_\Sigma\,,
$$
where $A$ can be any tensorial quantity defined in both regions.}
\begin{equation}
    [g_{\alpha\beta}] = 0\,,\;
    [K_{\alpha\beta}] = 0\,,\;
    [R] = 0\,,\;
    [\nabla_\alpha R] = 0\,.
    \label{junction_conditions_fR}
\end{equation}
In the case of a spherical static star, Eqs. \eqref{junction_conditions_fR} translate into
\begin{equation}
\left\{
\begin{aligned}
    A_\text{ext}(r_*)=A_\text{int}(r_*)\\
    A'_\text{ext}(r_*)=A'_\text{int}(r_*)\\
    B_\text{ext}(r_*)=B_\text{int}(r_*)\\
    R_\text{ext}(r_*)=R_\text{int}(r_*)\\
    R'_\text{ext}(r_*)=R'_\text{int}(r_*)
\end{aligned}
\quad,\right.
\end{equation}
where $r_*$ is the stellar radius.
Also, due to the last two equalities in  \eqref{junction_conditions_fR}, it also must be the case that 
\begin{equation}\label{eq:junction_noshell_5}
    n^\alpha[T_{\alpha\beta}]=0\,,
\end{equation}
where $n^\alpha$ represents a vector orthogonal to the hypersurface (the surface of the star) $\Sigma$. In the case of a spherical static star endowed with standard matter, modelled as a perfect fluid, the last condition yields
\begin{equation}
\label{matter_rho_P_at_rb}
P(r_*)=\rho(r_*)=0\,,
\end{equation}
which serves to compute the stellar radius.

Taking these into account and imposing asymptotic flatness (meaning that the metric coefficients should tend to their respective Minkowski values for large radii)
\begin{equation}
    \lim_{r\to\infty}A(r)=1
    \label{eq:A_asintotico_Sch}
\end{equation}
via the shooting method explained in \cite{AparicioResco:2016xcm}, one gets the initial value $A(0)$ that yields a physical solution. For computational reasons, the initial value of the scalar curvature is taken to be the same as in GR ($R(0)=\kappa T(0)$) as a suitable approximation.

In this work, we adopt the quadratic Starobinsky model, defined by $f(R) = a R^2$, with the parameter $a<0$ always in km$^2$. This choice is motivated by its simplicity, its well-studied role in early-universe inflation, and the fact that it introduces minimal yet nontrivial deviations from General Relativity. Moreover, the $R^2$ term captures leading-order corrections expected from quantum gravity effects and serves as a useful benchmark for studying modified gravity in the strong-field regime emerging both inside and in the neighborhood of compact objects \cite{Donoghue:1994dn}. We focus on the $a<0$ branch because it allows us to find consistent numerical solutions,  whereas we were unable to obtain asymptotically well-behaved solutions in the 
$a>0$ branch. 

\begin{figure}
\begin{center}
    \centering
    \includegraphics[width=0.5\linewidth]{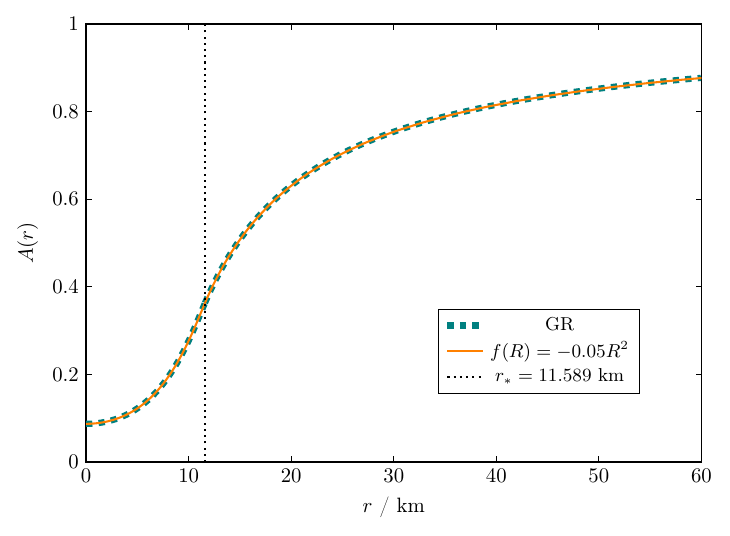}
    \hspace{-0.3cm}
    \includegraphics[width=0.5\linewidth]{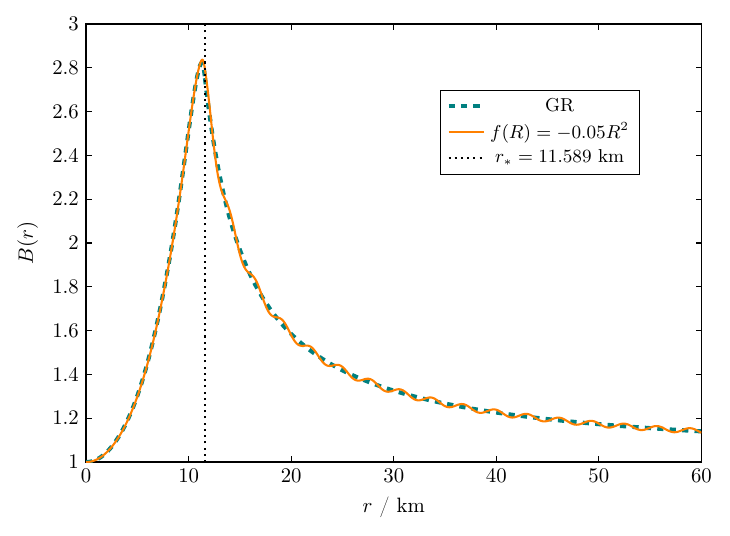}
    \includegraphics[width=0.53\linewidth]{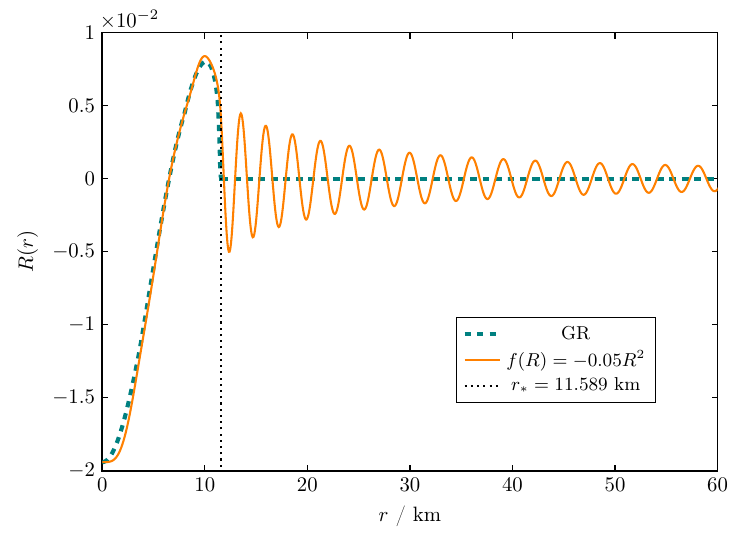}
    \caption{\footnotesize Metric coefficients $A(r)$ and $B(r)$, as well as the Ricci scalar curvature $R(r)$, compared for GR and $f(R)=-0.05R^2$, using the \textit{Middle} EoS for a star with central pressure $P_c=7\cdot10^{-4}$ km$^{-2}$ and a radius of 11.589 km. The main feature is an oscillation of the metric functions and the Ricci scalar outside the star around their GR counterparts, most visible on $R(r)$ (which is hence non-vanishing outside the star) and somewhat in $B(r)$.}
    \label{fig:ABR_comparados}
\end{center}
\end{figure}

After numerical integration, we obtain the interior and exterior glued metric coefficients---see Fig. \ref{fig:ABR_comparados}---as well as other functions such as the pressure or energy density inside the star. A major result is the non-vanishing scalar curvature in the outside vacuum region. This fact motivates an appropriate definition of a cumulative mass function, as discussed in \cite{rivero2024condiciones}. For illustrative purposes, these results, for the EoS labelled \textit{Middle} and a given central pressure are shown in Fig.  \ref{fig:ABR_comparados}, where a  comparison with those of GR is provided. The solutions were obtained via numerical integration of the Einstein field equations and careful application of the junction conditions established in \cite{Senovilla:2013vra}.

A characteristic feature of the solutions obtained is clearly visible in the form of a periodic oscillation, being more evident in the representation of the scalar curvature in Fig. \ref{fig:ABR_comparados}. Outside of the star, we observe a decaying amplitude with an oscillation period that can be well approximated in the middle and far region by an expression of the form $R(r)\sim  \frac{\cos(m_{\varphi}r+\delta_0)}{r}$. This behavior can be understood by taking the trace of Eq. (\ref{fieldEQS}),  which leads to 
\begin{equation}
    \Box R+\frac{1}{6a}R=\kappa T \ .
\end{equation}
In the far region (weak gravity and asymptotically Minkowskian background), this equation admits two types of solutions, depending on the sign of $a$. If $a>0$ we get a combination of positive and negative exponentials, ${\rm e}^{\pm m_\varphi r}/r$, while for $a<0$ we get, instead, complex solutions, which can be combined in the form $R(r)\sim  \frac{\cos(m_{\varphi}r+\delta_0)}{r}$ mentioned before, with $m_\varphi=1/6|a|$ (see Appendix \ref{AppendinxA} for more details). The existence of a positive growing exponential when $a>0$ explains the difficulties to find reasonable numerical solutions far from the star. In such regions, an extreme fine tuning would be necessary to prevent the ${\rm e}^{+ |m_\varphi| r}/r$ from switching on in the numerical analysis. On the contrary, the oscillating solutions can be implemented more straightforwardly, which justifies our choice of branch for $a$. 

Although we are mainly interested in the qualitative behavior of the solutions, we have numerically verified that the oscillation frequency far away is consistent with the expected value $m_\varphi=1/6|a|$, with an additional exponential damping of the amplitude which we assume is of numerical origin. Convergence tests to check the validity of our solutions have also been implemented, as briefly discussed in Appendix \ref{app:numcheck}.

\section{Massive particle orbits in static and spherically symmetric spacetimes in $f(R)$ theories}
\label{sec:massive orbits}

Having obtained in the previous section the realistic glued inner and outer metrics for neutron stars embedded in an $f(R)=aR^2$ theory, we may now use them in our exploration of orbital motions. In this context, one of the most relevant analyses relies on the study of massive particle trajectories in the vicinity of the star, which are instrumental to understanding the dynamics in the presence of strong gravitational fields. In particular, we are interested in the existence of stable circular orbits (SCOs) around the star, and whether they show different quantitative and qualitative behavior compared to their GR counterparts.

\subsection{Stable circular orbits of massive particles}
\label{SecIIIA}

First, because of the spherical symmetry of the metrics under study, we shall simplify the computations by only considering particles moving in the equatorial plane, i.e., $\theta=\pi/2$ without loss of generality.

For any static and spherically symmetric metric of the form \eqref{metric}, the Lagrangian has the general form 
\begin{equation}
\mathcal{L}=\left(A\dot{t}^2-B\dot{r}^2-r^2\dot{\phi}^2\right)^{1/2},
\label{Lagrangian}
\end{equation}
Since in the context of $f(R)$ theories the metric coefficients $A(r)$ and $B(r)$ are in general not analytic but numerical functions, the Lagrangian above cannot be a priori simplified further. However, the Euler-Lagrange equations for $t$ and $\phi$ result in
\begin{equation}
A\dot{t}=k\;,\;
    r^2\dot{\phi}=h\,.
    \label{EL-eqns}
\end{equation}
which we identify as the conservation of \textit{specific} (meaning \textit{per unit mass}) energy $k$ and angular momentum $h$ of the particle, respectively. Thus, upon substitution of these two conserved quantities in the expression of the Lagrangian \eqref{Lagrangian}
with $\mathcal{L}=\epsilon$, and defining $\xi=\frac{1}{r}$, one obtains\footnote{
where we have used the chain rule to obtain
\begin{equation}
    \frac{{\rm d} r}{{\rm d} \phi}=-\frac{1}{\xi^2}\frac {{\rm d}\xi}{{\rm d}\phi}.
\end{equation}
}
\begin{equation}
    \epsilon^2=\frac{k^2}{A}-Bh^2\left(\frac{{\rm d}\xi}{{\rm d}\phi}\right)^2-h^2\xi^2,
\label{eq:epsilon^2_f(R)}
\end{equation}

As is well known, the previous expression can be manipulated to obtain an energy conservation equation outside the star for either massive particles ($\epsilon=1$) or light ($\epsilon=0$). For the first case, identifying the terms corresponding to kinetic energy, effective potential, and total energy, one gets
\begin{equation}   
\frac{1}{2}AB\dot{r}^2+ V_{\rm eff}=\frac{1}{2}(k^2-1)\,,
\label{eq:energy_massive}
\end{equation}  
where $AB\neq1$, unlike in GR whose exterior is the usual Scwharzschild solution. The effective potential takes the form  
\begin{equation}  
\label{V_eff}  
V_\text{eff}=\frac{1}{2}A\left(1+\frac{h^2}{r^2}\right)-\frac{1}{2}.  
\end{equation}  
Next, we impose the conditions for circular orbits at an eventual $r=r_0$, in particular
\begin{eqnarray}
\dot{r}=0\,,\;\; \left.\frac{{\rm d}V_{\rm eff}}{{\rm d}r}\right|_{r_0}=0\,,
\label{circular_massive}
\end{eqnarray}
and their stability 
\begin{eqnarray}
 \left.\frac{{\rm d}^2 V_{\rm eff}}{{\rm d}r^2}\right|_{r_0}\geq0\,. 
 \label{stable_massive}
\end{eqnarray}
The first condition yields a relationship between the three relevant variables $k$, $h$ and $r_0$:  
\begin{equation}  
    k^2=A(r_0)\left(\frac{h^2}{r_0^2}+1\right). 
    \label{eq:k^2}
\end{equation}  
After some manipulation, the second condition provides an expression for $h$ as a function of the orbital radius: %
\begin{equation}  
    h^2=\frac{r_0^3}{2\left.\frac{A}{A'}\right|_{r_0}-r_0},  
    \label{eq:h^2}
\end{equation}  
which, upon substitution into the previous equation, leads to $k=k(r_0)$:  
\begin{equation}  
    k^2=\frac{A(r_0)}{1-\frac{1}{2}r_0
    \left.\frac{A}{A'}\right|_{r_0}}.  
    \label{eq:k^2}
\end{equation}  
With these two expressions, for a given orbital radius $r_0$, one can determine the values of $h$ and $k$ that lead to bound and circular orbits, provided certain existence conditions are met. Specifically, for the orbits to be physical, both $k$ and $h$ must be real numbers, so
\begin{equation}  
    k^2>0,\quad h^2>0,
\end{equation}  
and for them to be bound, the total energy from Eq. \eqref{eq:energy_massive} must be negative, so
\begin{equation}
    k^2<1.
\end{equation}
Furthermore, for the orbits to be stable---the third condition previously stated---Eq. \eqref{stable_massive} must also be satisfied, leading to  
\begin{equation}  
 \left(A''-2\frac{A'^2}{A}+3\frac{A'}{r}\right)\left(\frac{1}{1-\frac12r\frac{A'}{A}}\right)\geq0.  
\end{equation}  
These three inequalities can be rewritten in terms of the orbital radial coordinate ($r>r_*$) as  
\begin{equation}
\left\{
\begin{aligned}
    &C_1(r)\equiv A'>0\,,\\
    &C_2(r)\equiv 1-A-\frac{r A'}{2A}>0\,,\\
&C_3(r)\equiv A''+\frac{3A'}{r}-\frac{2A'^2}{A}>0\,,
\end{aligned}
\quad,\right.
\label{eq:SCOs_conditions}
\end{equation}
which allows us to determine the orbital radii $r_0$ for which stable circular orbits for massive particles exist.\\

\begin{figure}
\begin{center}
    \centering
    \includegraphics[width=0.8\linewidth]{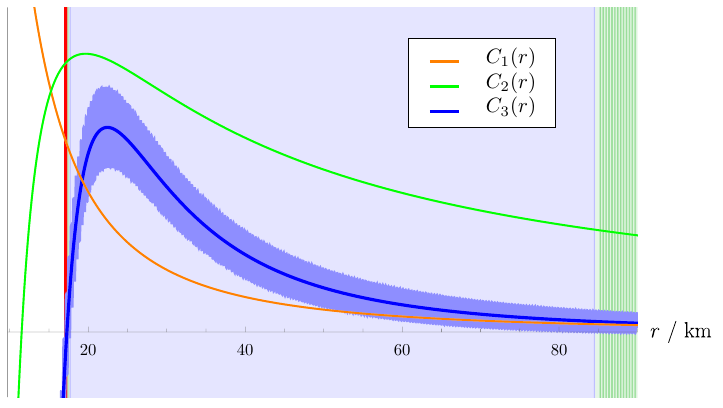}
    \caption{\footnotesize Radial condition functions $C_{i=1,2,3}(r)$, along with the SCO existence rings they produce (light green regions) for the \textit{Soft} EoS and a central pressure $P_c=6\cdot10^{-4}$ km$^{-2}$, both for GR (darker) and $f(R)=-0.00062R^2$ (lighter and oscillating). The oscillations are only perceptible for $C_3^{f(R)}(r)$ in the depicted scales. %
    The principal ring is highlighted in light violet, and the ISCO is illustrated as a red vertical line at $r^\text{GR}_\text{ISCO}=6M=17,273$ km. 
    The left border of the figure corresponds to the stellar radius $r_*=9,731$ km, equal for both GR and $f(R)$. Since only their signs concern us, all condition functions have been rescaled by adequate positive constants to ensure visual comparability, without any  loss of validity.
    }
    \label{fig:condRG}
\end{center}
\end{figure}

\textbf{* GR case:} To study these regions, we first ana\-lyze the behavior of the conditions $C_{i=1,2,3}$ in GR as per \eqref{eq:SCOs_conditions}. Indeed, by substituting the expression of the coefficient $A$ for the exterior Schwarzschild metric into these conditions, we obtain %
\begin{equation}
\left\{
\begin{aligned}
    &C_1^{\rm GR}(r)=\frac{2M}{r^2}>0\,,\\
    &C_2^{\rm GR}(r)=\frac{M}{r}\left(\frac{r-4M}{r-2M}\right)>0\,,\\
    &C_3^{\rm GR}(r)=\frac{1}{r^{3}}\left(2M-\frac{8M^{2}}{r-2M}\right)>0\,,
\end{aligned}
\right.
\label{eq:SCOs_GR_conditions}
\end{equation}
which are represented in Fig. \ref{fig:condRG}. The functions $C_2^{\rm GR}$ and $C_3^{\rm GR}$ tend to $-\infty$ at the Schwarzschild radius $r_s\equiv2M$ and cross zero at $2r_s\equiv4M$ and $3r_s\equiv6M$, respectively, while $C_1^{\rm GR}$, i.e. $A'(r)$, remains positive throughout its domain (the function $A(r)$ in the Schwarzschild metric is strictly increasing). Thus, in the GR context, all three conditions exhibit a horizontal asymptote at zero from the positive side, implying that $C_{i=1,2,3}^{\rm GR}>0$ for all $r>6M$, so that stable circular orbits exist beyond this value of the radial coordinate, denoted as $r^{\rm GR}_{\rm ISCO}\equiv6M$, widely known as the \textit{Innermost Stable Circular Orbit} (ISCO).\\

\begin{figure*}[t]
\begin{center} 
    \centering
    \includegraphics[width=0.4\linewidth]{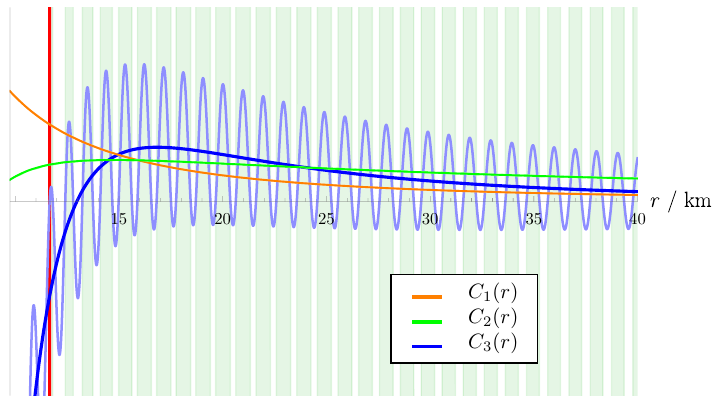}\hspace{-0.075cm}
    \includegraphics[width=0.4\linewidth]{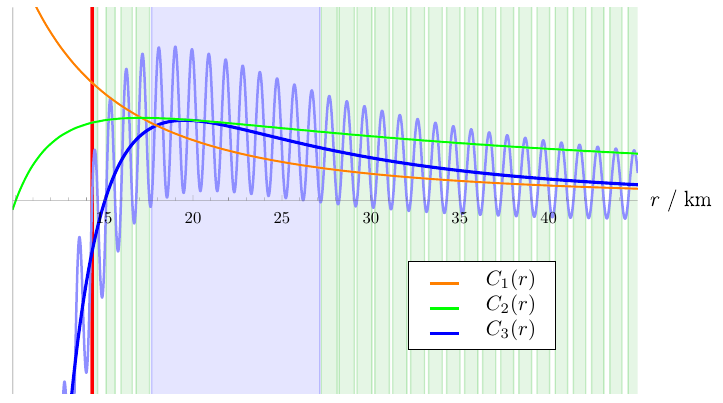}
    \includegraphics[width=0.4\linewidth]{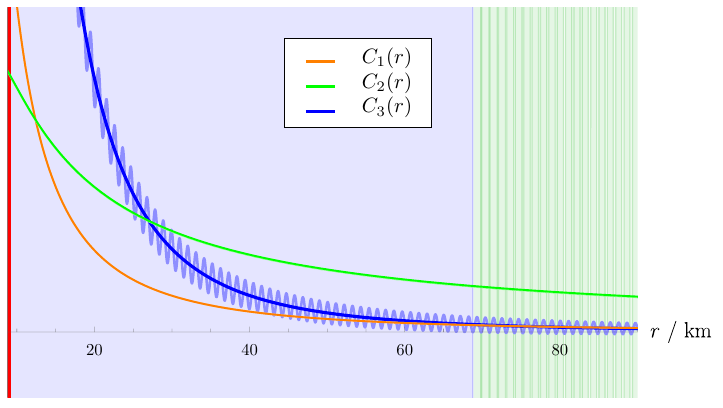}
    \includegraphics[width=0.4\linewidth]{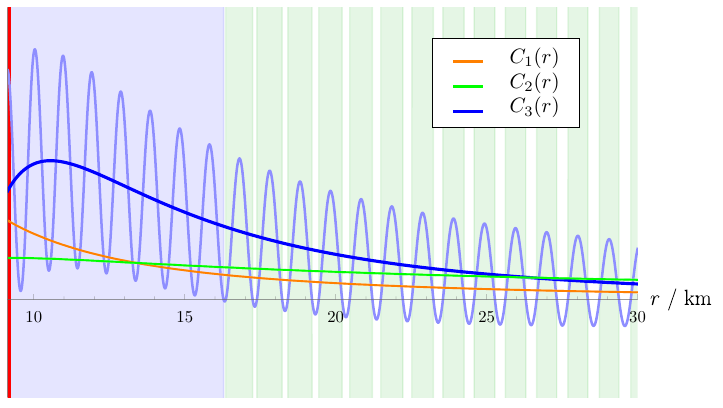}
    \caption{\footnotesize Radial condition functions $C_{i=1,2,3}(r)$, along with the SCO existence rings they produce (light green regions) for the \textit{Soft} EoS, both for GR (darker) and $f(R)=-0.005R^2$ (lighter and oscillating). The oscillations are only perceptible for $C_3^{f(R)}(r)$ at this scale. The principal ring is highlighted in light violet, and the ISCO is illustrated as a red vertical line.
 The vertical red and grey lines on the left of each panel correspond to the ISCOs and the stellar radii respectively. \textbf{Upper left:} $P_c=2\cdot10^{-4}$ km$^{-2}$, $r_*=9,742$ km, $r^\text{GR}_\text{ISCO}=12,972$ km, $r^{f(R)}_\text{ISCO}=11,649$ km; the oscillations on $C_3^{f(R)}(r)$ pass through zero repeatedly everywhere further than the ISCO, so there is no principal ring. \textbf{Upper right:} $P_c=3\cdot10^{-4}$ km$^{-2}$, $r_*=9,865$ km, $r^\text{GR}_\text{ISCO}=15,048$ km, $r^{f(R)}_\text{ISCO}=14,320$ km; $C_3^{f(R)}(r)$ is constantly positive in an intermediate region, resulting in a principal ring much wider than all the others. For higher values of $r$ rings continue appearing. \textbf{Lower left:} $P_c=6\cdot10^{-5}$ km$^{-2}$, $r_*=r^{f(R)}_\text{ISCO}=8,806$ km; the ISCO exactly coincides with the stellar radius and lies within the principal ring. \textbf{Lower right:} $P_c=9\cdot10^{-5}$ km$^{-2}$, $r_*=r^{f(R)}_\text{ISCO}=9,132$ km; the stellar radius is small enough for the maximum of $C_3(r)$ to show.
}
    \label{fig:condiciones}
\end{center}
\end{figure*}

\textbf{* $\boldsymbol{f(R)}$ case:}
As described in the previous section, the metric coefficients 
ensuring smooth matching in the absence of thin shells or double layers turn out to be numerical functions of the radial coordinate. Consequently, the analysis of the existence and stability of circular orbits in this context for the exterior of static neutron stars must also be numerical. Given that the main effect of $f(R)$ theories on the exterior metric is the introduction of damped oscillations, their impact on the conditions $C_{i=1,2,3}$ will also be of the same nature, as shown in Figs.  \ref{fig:condRG} and  \ref{fig:condiciones}.

\begin{figure*}[htbp!]
\begin{center}
    \centering
    \includegraphics[width=0.3\linewidth]{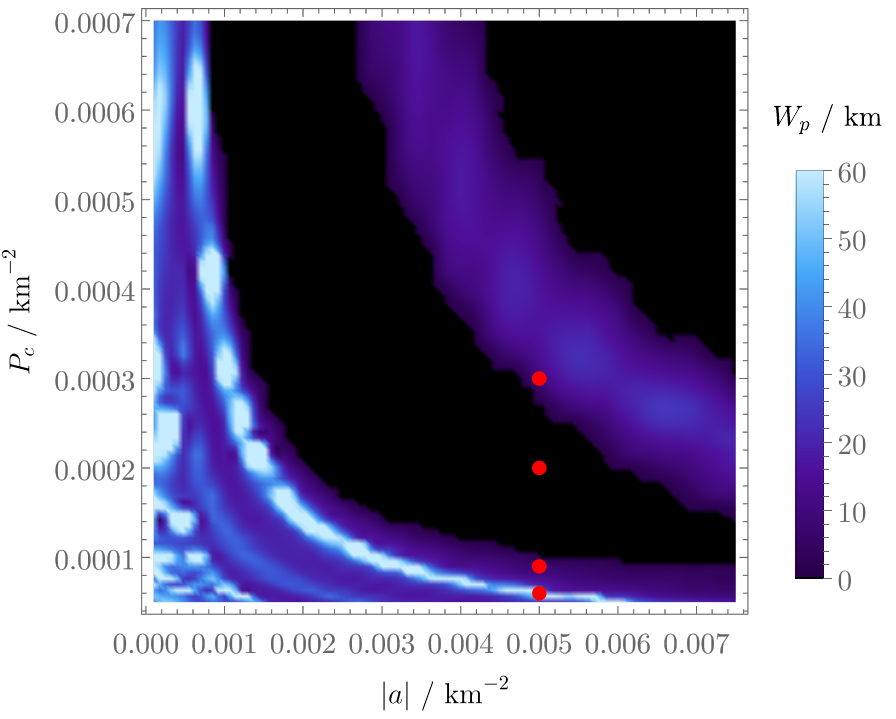}
    \includegraphics[width=0.3\linewidth]{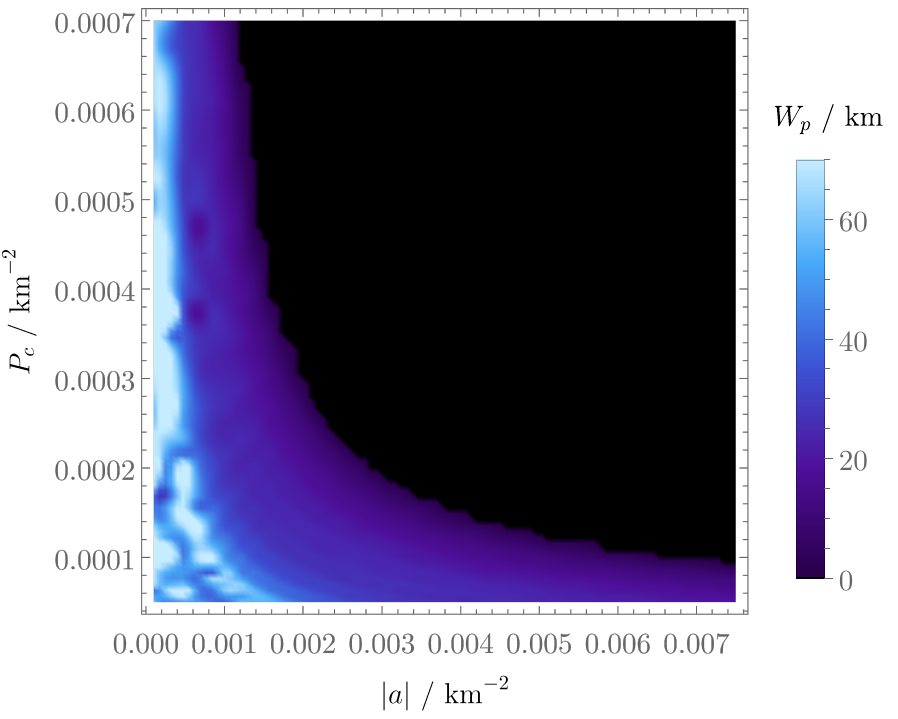}
    \includegraphics[width=0.3\linewidth]{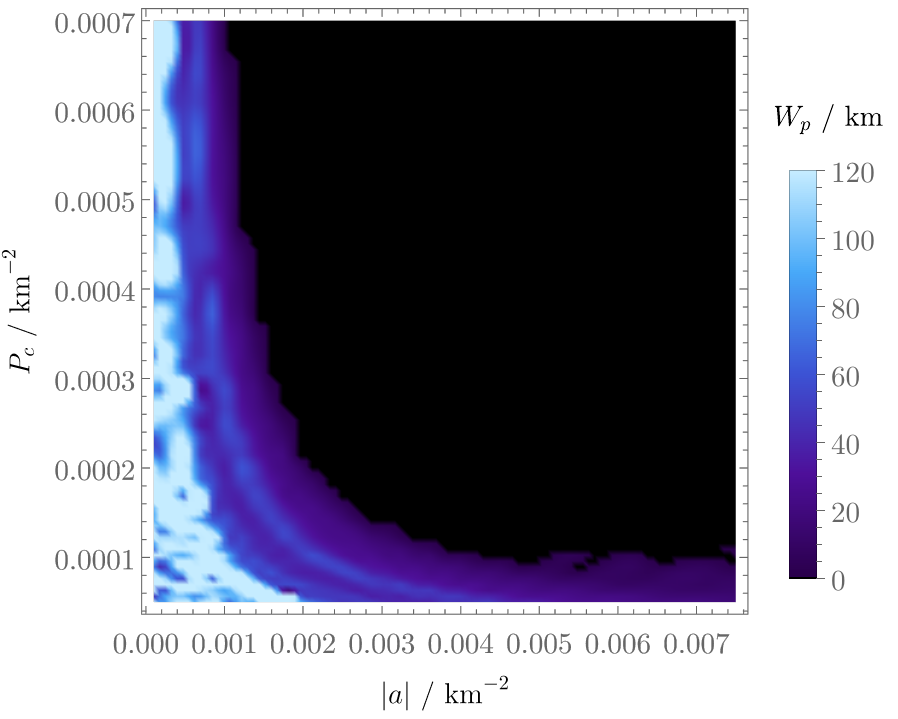}
    \includegraphics[width=0.3\linewidth]{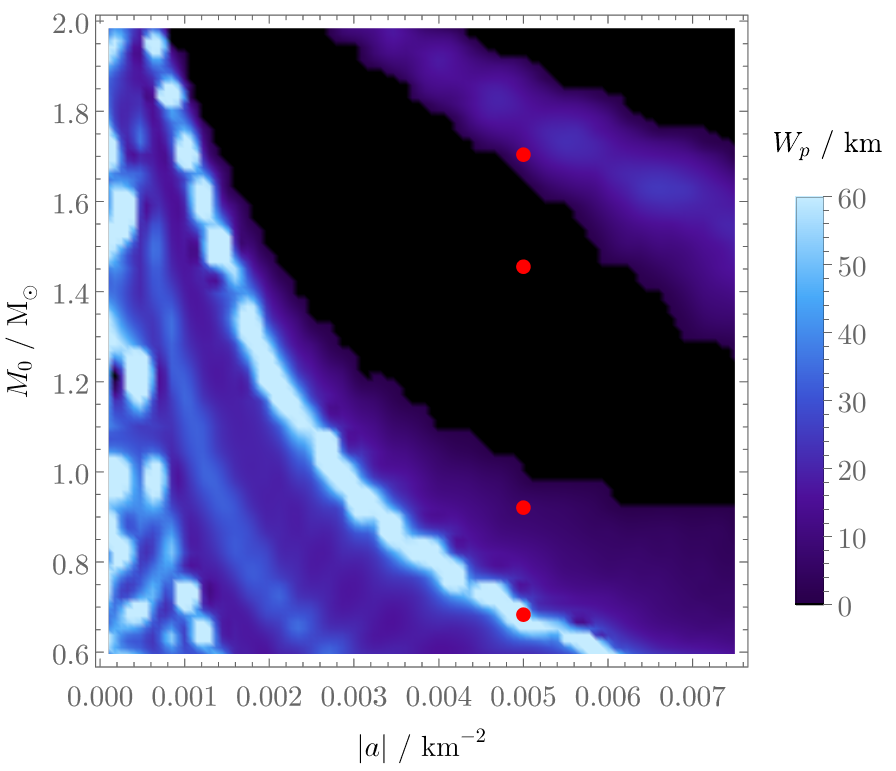}
    \includegraphics[width=0.3\linewidth]{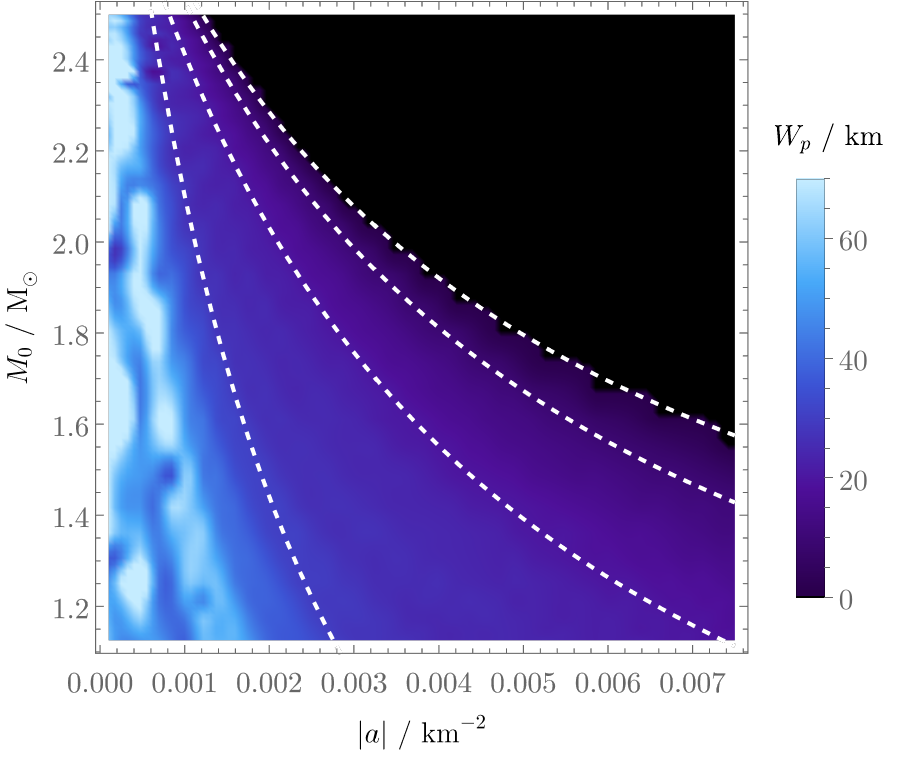}
    \includegraphics[width=0.3\linewidth]{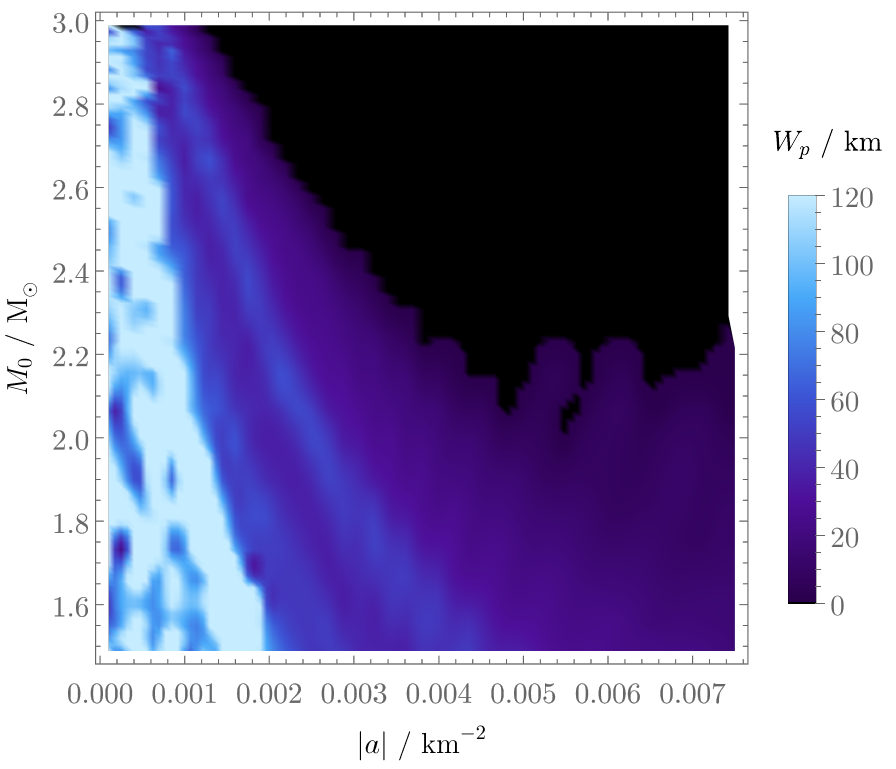}
    \caption{\footnotesize Principal ring width $W_p$ versus the coefficient $|a|$ and central pressure $P_c$ / bare stellar mass $M_0$ for the EoS \textit{Soft}, \textit{Middle} and \textit{Stiff}, respectively. The areas in black correspond to an absence of principal ring. A maximum color gradient limit has been set for the ring width representation, in order to avoid excessive distortion of the gradient and ensure that variations within the lower width range are clearly perceptible. Marked in red are the points studied in Fig. \ref{fig:condiciones}. Contour lines for $W_p=1,10,20,30$ km are shown for the EoS \textit{Middle}.}
    \label{fig:wp}
\end{center}
\end{figure*}
In all the depicted cases in the aforementioned figures, and in general for the range of central pressures, values of $a$ and EoS studied in this work, we have concluded that given the low amplitude of the oscillations in $C_1^{f(R)}(r)$ and $C_2^{f(R)}(r)$, both remain positive in the vicinity of the star, meaning that the determining condition is $C_3^{f(R)}(r)$. Thus, we can categorize the $C_3^{f(R)}(r)$ behavior into four regions:
\begin{itemize}
    \item $r_*<r<r_{\rm ISCO}^{f(R)}$: $C_3^{f(R)}(r)$ stays negative, so no SCOs exist in this region. At $r=r_{\rm ISCO}^{f(R)}$, $C_3^{f(R)}(r)$ vanishes by definition.
    \item $r_{\rm ISCO}^{f(R)}<r<r_{p_-}$: $C_3^{f(R)}(r)$ oscillates crossing zero, leading to narrow \textit{rings of solutions}.
    \item $r_{p_-}<r<r_{p_+}$: $C_3^{f(R)}(r)$ is positive, creating a wide region of SCO existence that we may call the \textit{principal ring}. 
    \item $r_{p_+}<r<\infty$: As $C_3(r)$ asymptotically approaches zero, its oscillations start crossing it again, producing narrow rings similar to those in the second region. Beyond a certain point, $C_1(r)$ and $C_2(r)$ also oscillate around a zero value as $r\rightarrow\infty$, 
    making the distribution and width of the rings difficult to describe phenomenologically.
\end{itemize}

\subsection{Width of the principal ring}
\label{SecIIIB}

The calculation of the principal ring width $W_p=r_{p+} - r_{p-}$ is performed numerically. 
Taking into account the 
condition functions $C_{i=1,2,3}^{f(R)}(r)$ it is possible to determine the different regions where the aforementioned functions are all positive and hence SCOs are allowed.
If a {\it principal} ring exists, it will correspond to the widest one.\footnote{To determine if the widest ring is indeed produced by some minimum of the oscillating $C_3(r)$ occurring over 0 and not just one period of the oscillation, any possible candidate to become a principal ring with $W_p<2\lambda_{C_3}$ is discarded, where $\lambda_{C_3}$ is the average wavelength of $C_3(r)$} Obviously, the principal ring width will be influenced by the amplitude of oscillations in $A(r)$ which in turn depend upon both the star's bare mass $M_0$, i.e., the $f(R)$ mass at the star's surface, as explained in \cite{rivero2024condiciones}, and its radius $r_*$. These two quantities are, after solving the system \eqref{eq:de_A_f(R)}-\eqref{eq:de_P_f(R)},  determined by the central pressure $P_c$, the coefficient $a$ that regulates the $f(R)$ model, and finally the specific neutron star's EoS. 
For illustrative purposes, comparative results between GR and a realization of the quadratic Starobinsky model are shown in Figs. 
\ref{fig:condRG} 
and  \ref{fig:condiciones}. 
In the latter, several values of central pressures have been considered to illustrate how both the ISCO and the very existence and main features of the principal ring crucially depend upon the central pressure. Also, a more  systematic analysis 
was performed in Fig. \ref{fig:wp} for all three EoS under consideration. Therein we considered a wide range of values of the central pressure and the $f(R)$ model parameter $a$. Accordingly, we concluded that the larger either $\lvert a\rvert$ or the central pressure are, the narrower the principal ring becomes, eventually disappearing. In the limit $a\to0$ the principal ring widens infinitely ($W_p\to\infty$), recovering the GR behaviour.

\subsection{Behavior of the ISCO}
\label{SecIIIC}
\begin{figure*}[ht!]
\begin{center}
    \centering
    \includegraphics[width=0.3\linewidth]{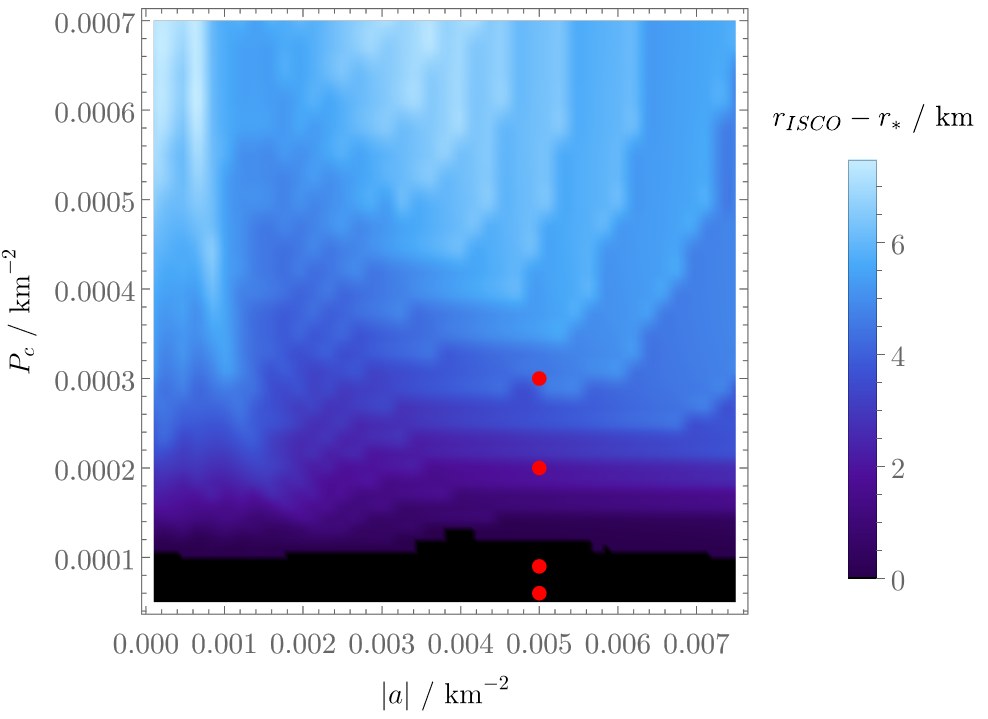}
    \includegraphics[width=0.3\linewidth]{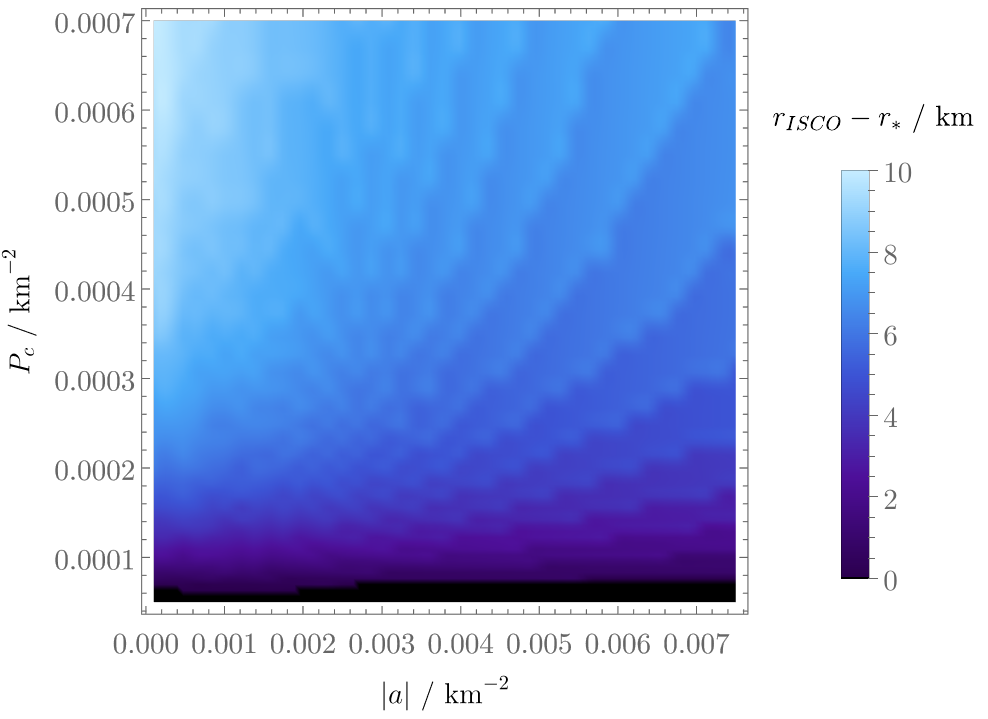}
    \includegraphics[width=0.3\linewidth]{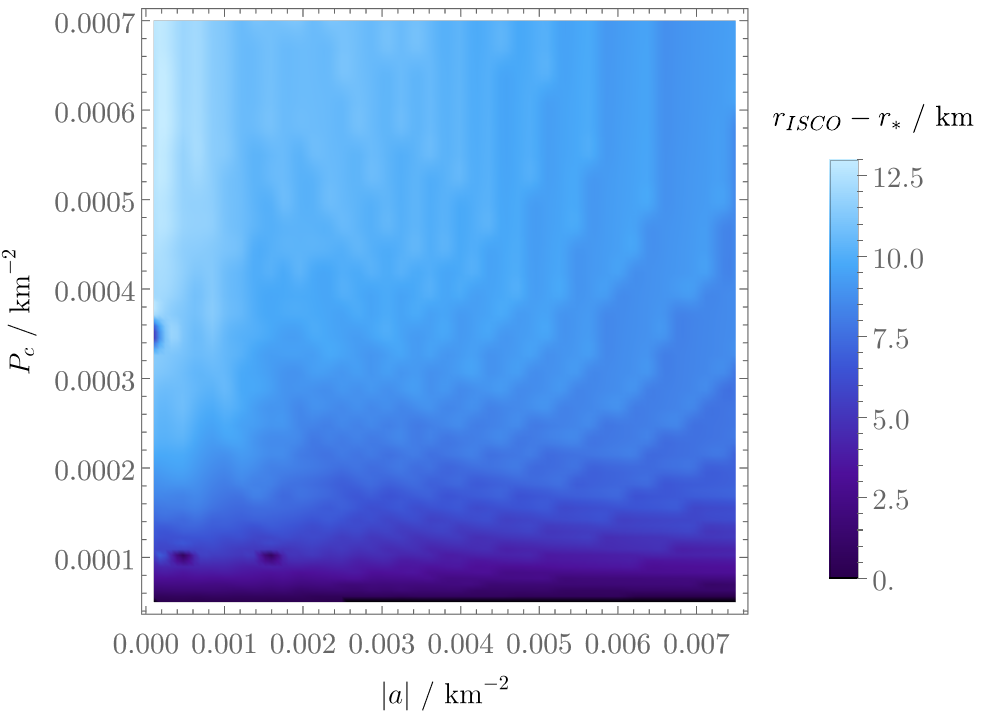}
    \includegraphics[width=0.3\linewidth]{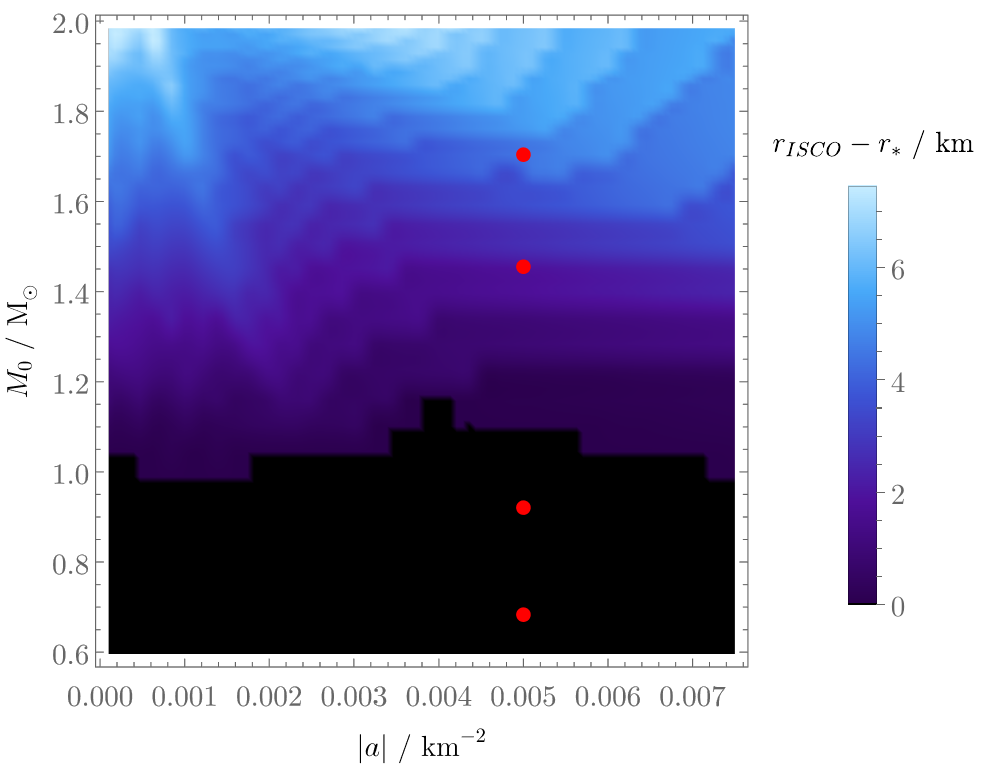}
    \includegraphics[width=0.3\linewidth]{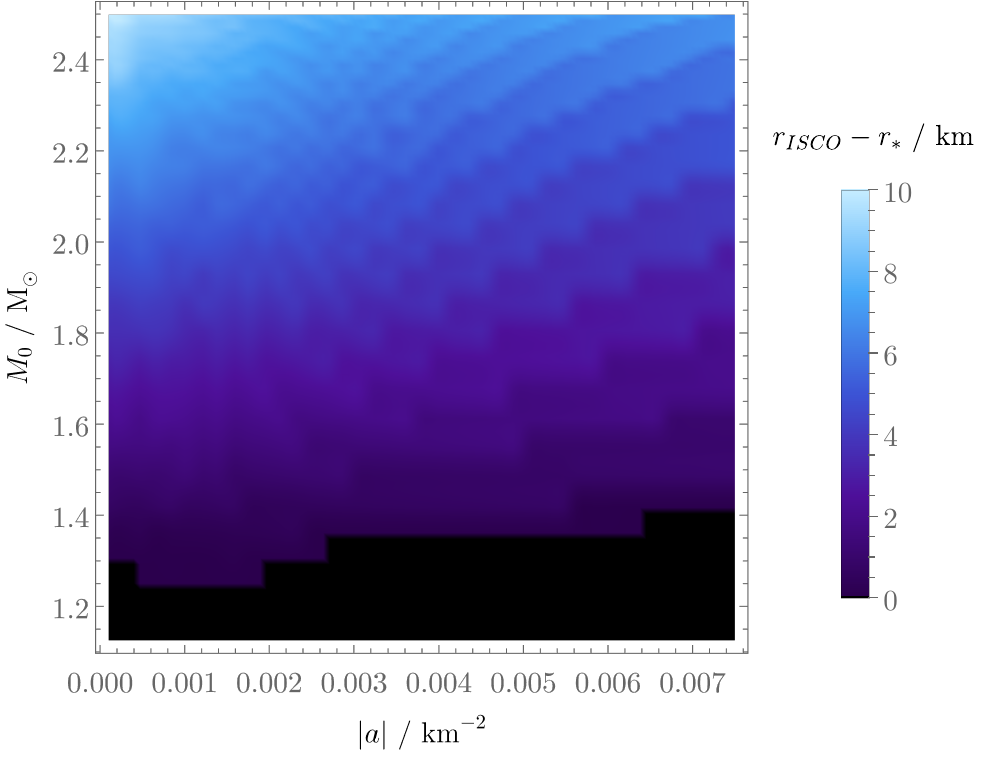}
    \includegraphics[width=0.3\linewidth]{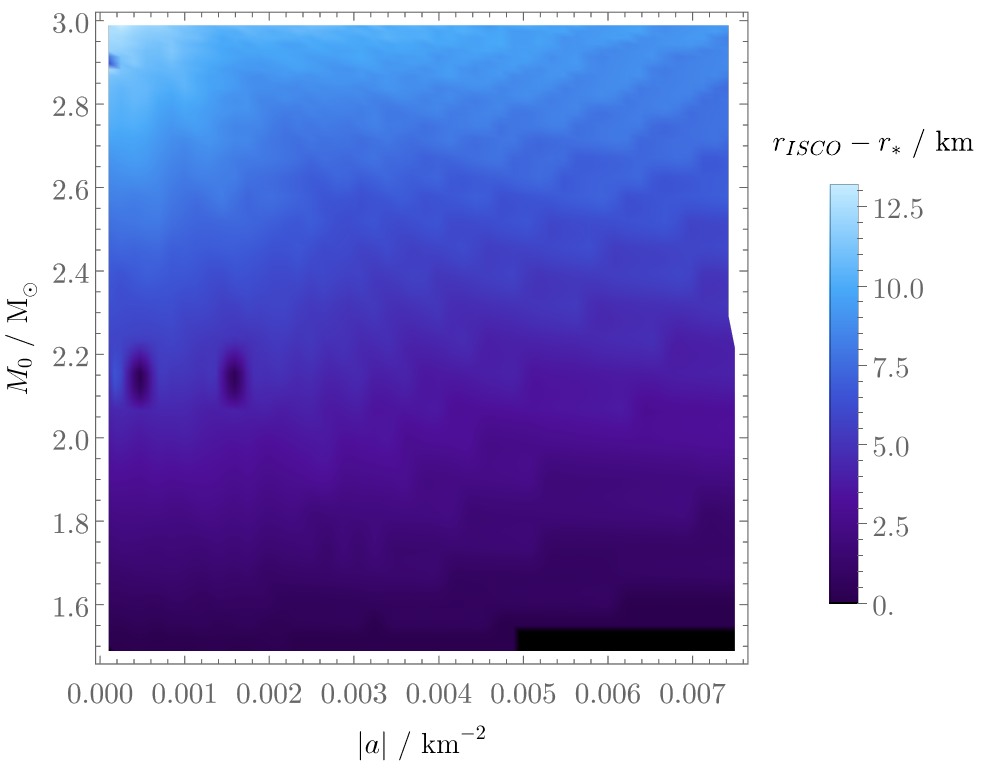}
    \caption{\footnotesize Distance between the ISCO radius $r_{ISCO}$ and the stellar radius $r_*$ versus the coefficient $|a|$ and central pressure $P_c$ / bare stellar mass $M_0$ for the EoS \textit{Soft}, \textit{Middle} and \textit{Stiff}, respectively. The areas in black correspond to the existence of SCOs arbitrarily close to the star's surface. Marked in red are the points studied in Fig. \ref{fig:condiciones}.}
    \label{fig:isco}
\end{center}
\end{figure*}

As it already happens in GR, the ISCO will sometimes effectively coincide with the stellar radius---resulting in stable circular orbits arbitrarily close to the surface---and other times the ISCO will lie outside it at some distance, giving rise to an inner region ($r_*<r<r_{\rm ISCO}$) devoid of SCOs. This behaviour is illustrated in Fig. \ref{fig:isco}.

\begin{figure*}[ht!]
\begin{center}
    \centering
    \includegraphics[width=0.3\linewidth]{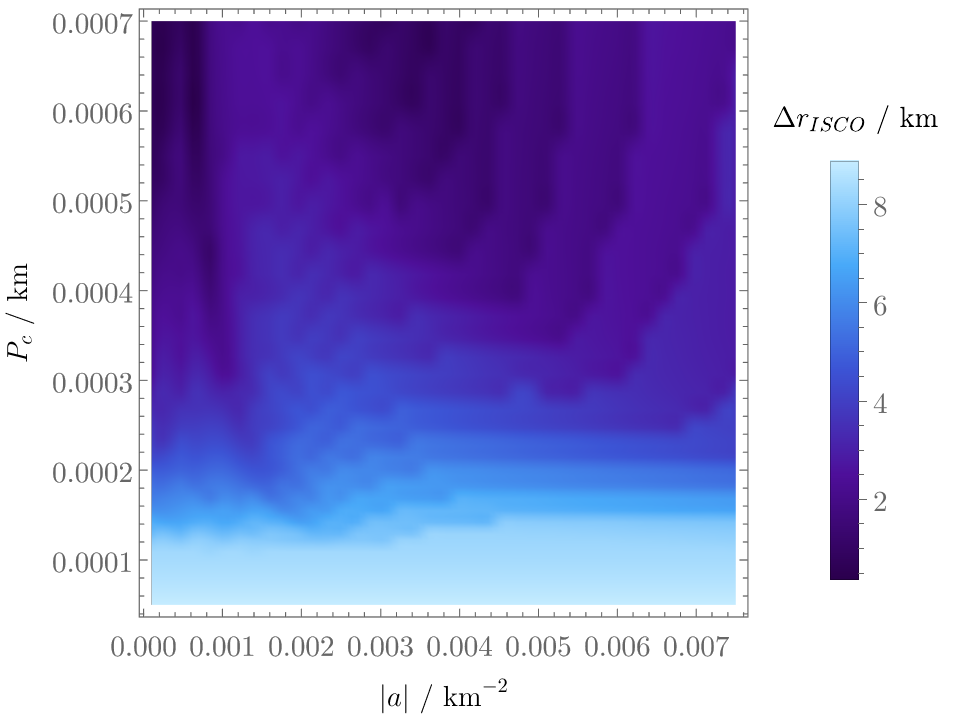}
    \includegraphics[width=0.3\linewidth]{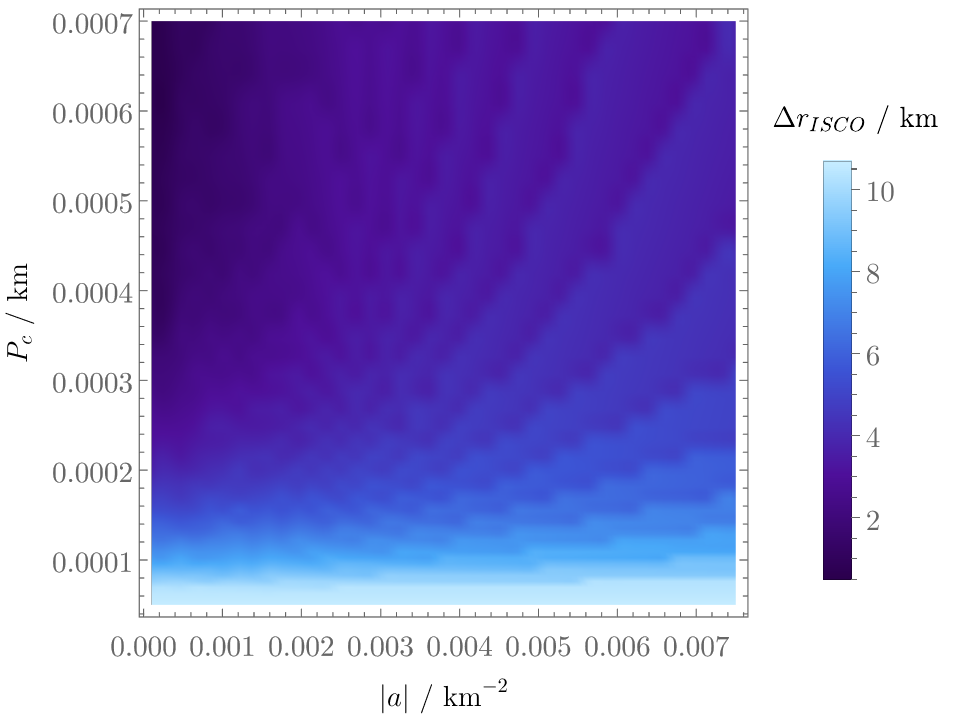}
    \includegraphics[width=0.3\linewidth]{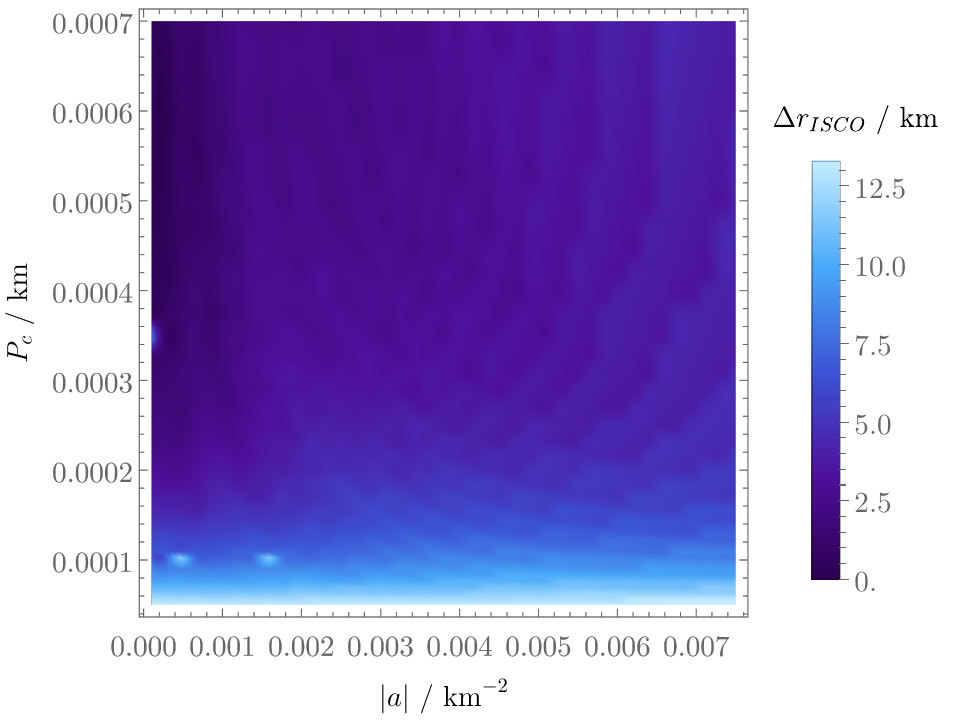}
    \includegraphics[width=0.3\linewidth]{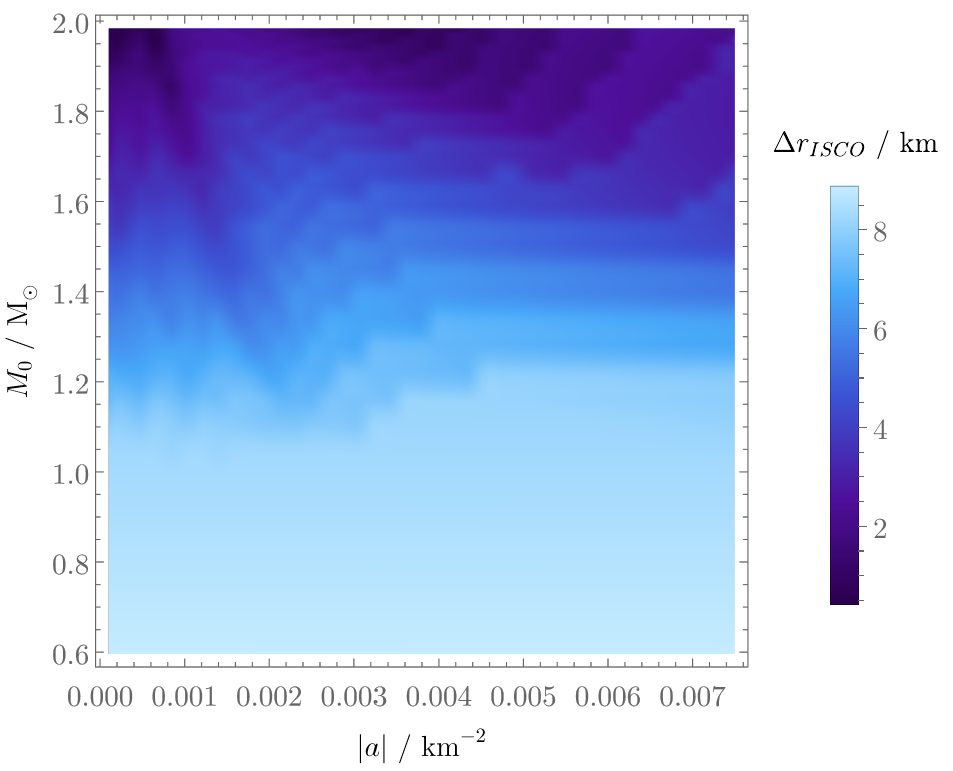}
    \includegraphics[width=0.3\linewidth]{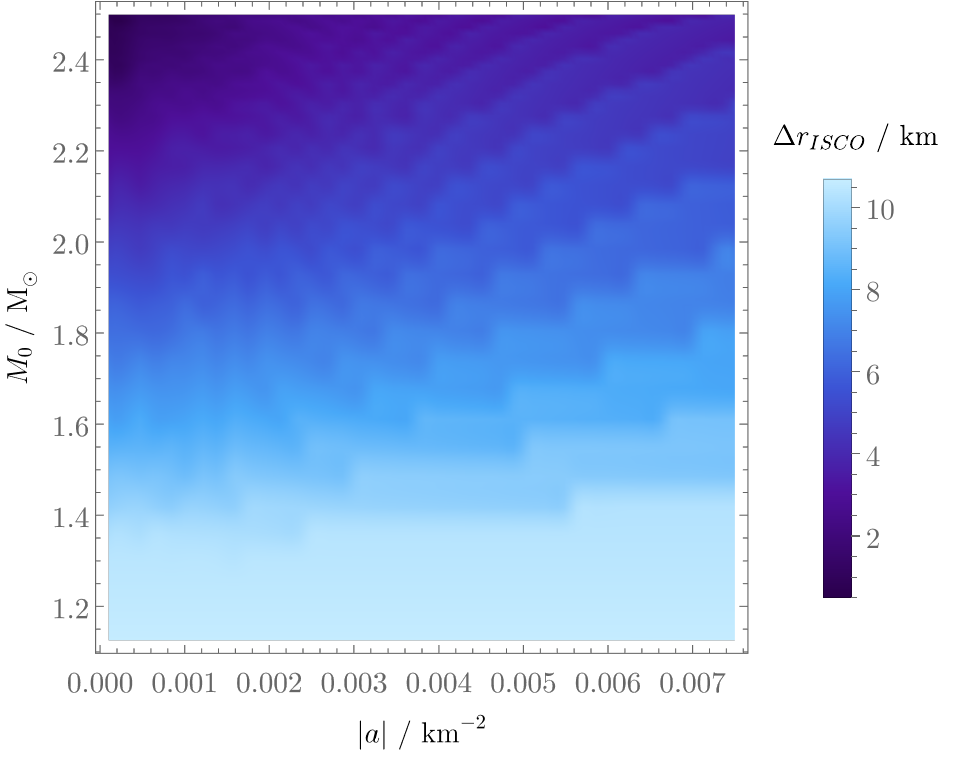}
    \includegraphics[width=0.3\linewidth]{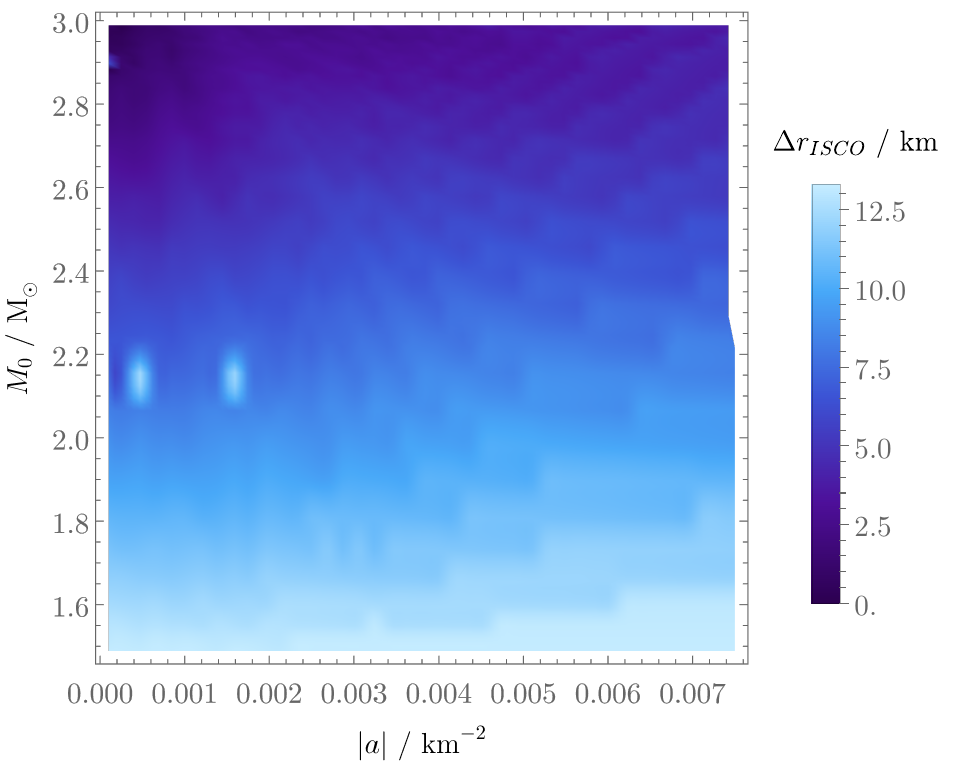}
    \caption{\footnotesize Difference between the ISCO radius $r_{\rm ISCO}$ in $f(R)$ and GR versus the coefficient $|a|$ and central pressure $P_c$ / bare stellar mass $M_0$ for the EoS \textit{Soft}, \textit{Middle} and \textit{Stiff}, respectively.}
    \label{fig:iscoRG}
\end{center}
\end{figure*}

\begin{figure}[htbp!]
\begin{center}
    \centering
    \includegraphics[width=0.8\linewidth]{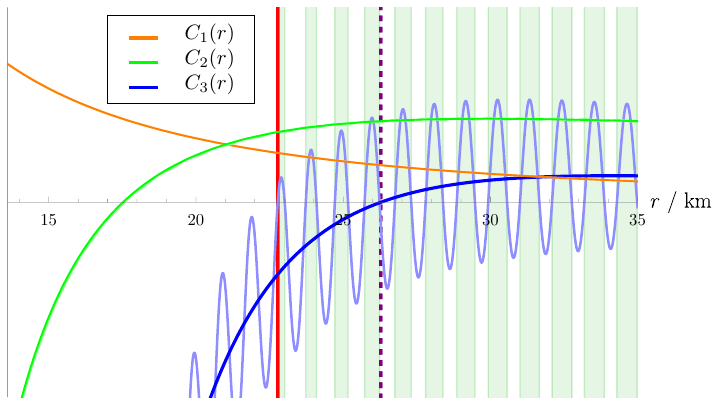}
    \caption{\footnotesize 
    Radial condition functions $C_{i=1,2,3}(r)$, along with the SCO existence rings they produce (light green regions) for the EoS \textit{Stiff} and a central pressure $P_c=5\cdot10^{-4}$ km$^{-2}$, both for GR (darker) and $f(R)=-0.007R^2$ (lighter and oscillating). For the latter, the stellar radius is depicted as the leftmost vertical grey line.  
    The oscillations are only perceptible for $C_3^{f(R)}(r)$ at this scale. The value $r_{\rm ISCO}^{f(R)}=22.778$ km is illustrated as a red vertical line, and the purple dashed line corresponds to the ISCO in GR, $r_{\rm ISCO}^\text{GR}=6M=26.278$ km, where $C_3^\text{GR}(r)$ turns positive.}
    \label{fig:iscos_RG_fR}
\end{center}
\end{figure}

Moreover, the ISCO in $f(R)$ theories can lie either inside or outside its GR counterpart. This is due to the fact that the stellar radius remains practically unaffected by variations in the parameter $a$, and is therefore identical for both GR and $f(R)$ scenarios. A comparative example of both cases is provided in Fig. \ref{fig:iscoRG}, and a specific example is shown in Fig. \ref{fig:iscos_RG_fR}. 

\subsection{Behavior of massive particles outside the SCOs}
\label{SecIIID}

The conditions \eqref{eq:SCOs_conditions} may be interpreted further as main attributes of the particle orbits. Namely, $C_1(r_0)$ corresponds to the existence of orbits with a turning point at $r_0$ (from now on, the \textit{turning radius}), $C_2(r_0)$ to their boundedness and $C_3(r_0)$ to their stability. As such, different combinations of these conditions holding or not give rise to different orbits, all with a turning point (or a point of pure angular motion) at the specified radius $r_0$, forced by construction when we fix $h$ and $k$ in  \eqref{eq:h^2} and \eqref{eq:k^2}, respectively.

\begin{figure}
\begin{center}
    \centering
    \includegraphics[width=0.8\linewidth]{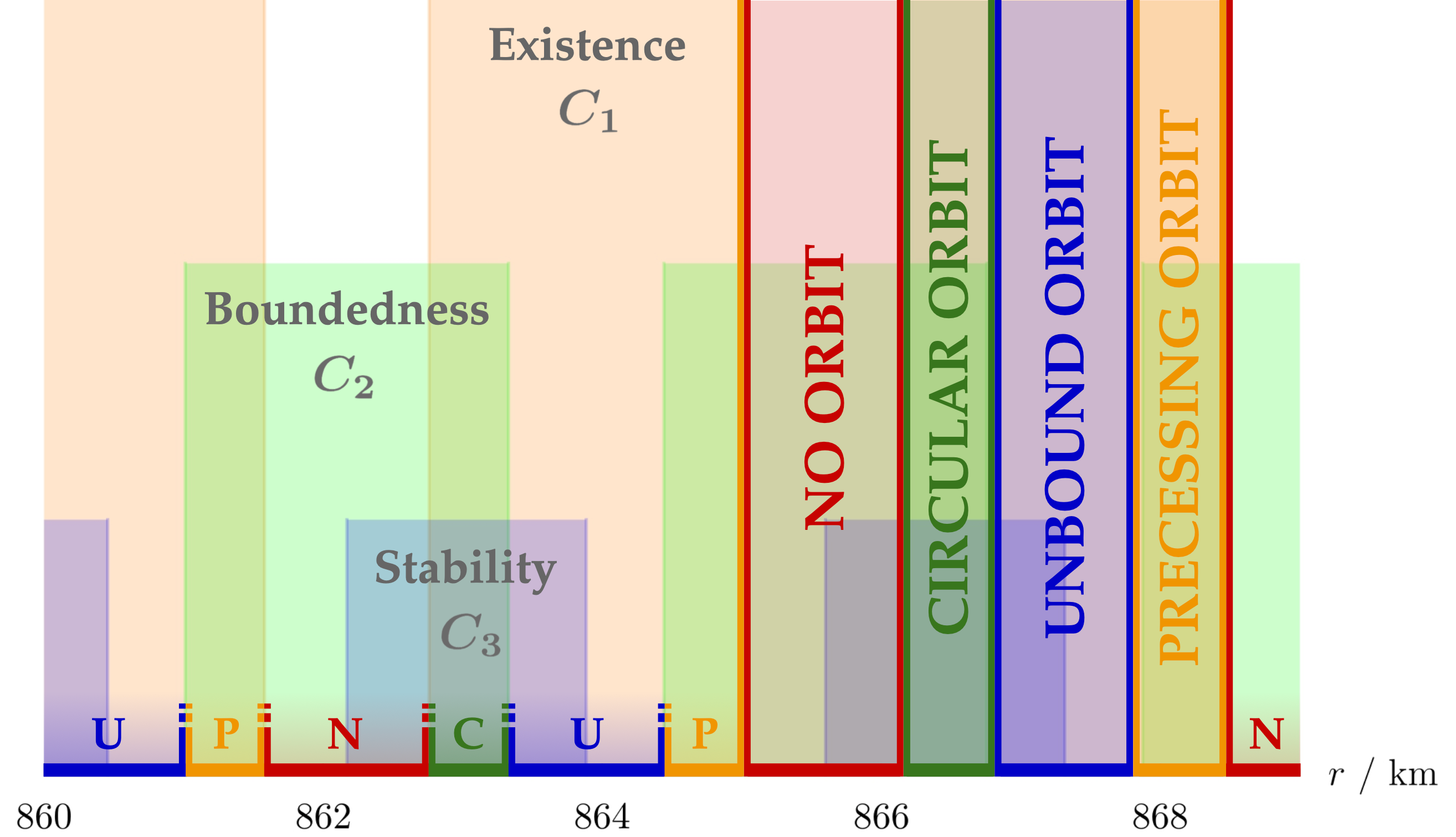}
    \caption{\footnotesize Conditions $C_i$ and combined effect on the resulting orbit for $f(R)=-0.05R^2$, $P_c=7\cdot10^{-4}$ km$^{-2}$ and the EoS \textit{Soft}. Since they depend on $A(r)$, an oscillating function, the conditions are quasi-periodic, and so are the resulting behavior rings. An example of an orbit for each of the highlighted rings is depicted in Fig. \ref{fig:orbitas}.}
    \label{fig:anilloscond}
\end{center}
\end{figure}

\begin{figure}
\begin{center}
    \centering
    \includegraphics[width=0.7\linewidth]{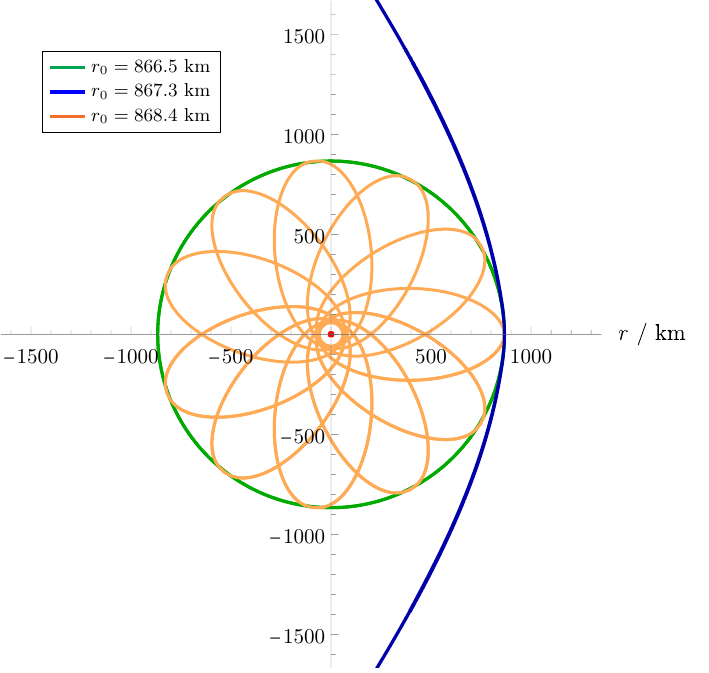}
    \caption{\footnotesize Orbits with turning points located at different orbit rings for $f(R)=-0.05R^2$, $P_c=7\cdot10^{-4}$ km$^{-2}$ and the EoS \textit{Soft}. In green, a circular orbit, satisfying the existence ($C_1$), boundedness ($C_2$) and stability ($C_3$) conditions. In orange, a precessing elliptical ($e=0.86$) orbit satisfying only the first two but failing to be stable, hence its non constant radius. In blue, a hyperbolic orbit ($e=2.09$) failing to satisfy the boundedness condition $C_2$, and hence escaping to infinity.}
    \label{fig:orbitas}
\end{center}
\end{figure}

\begin{figure}
\begin{center}
    \centering
    \includegraphics[width=0.7\linewidth]{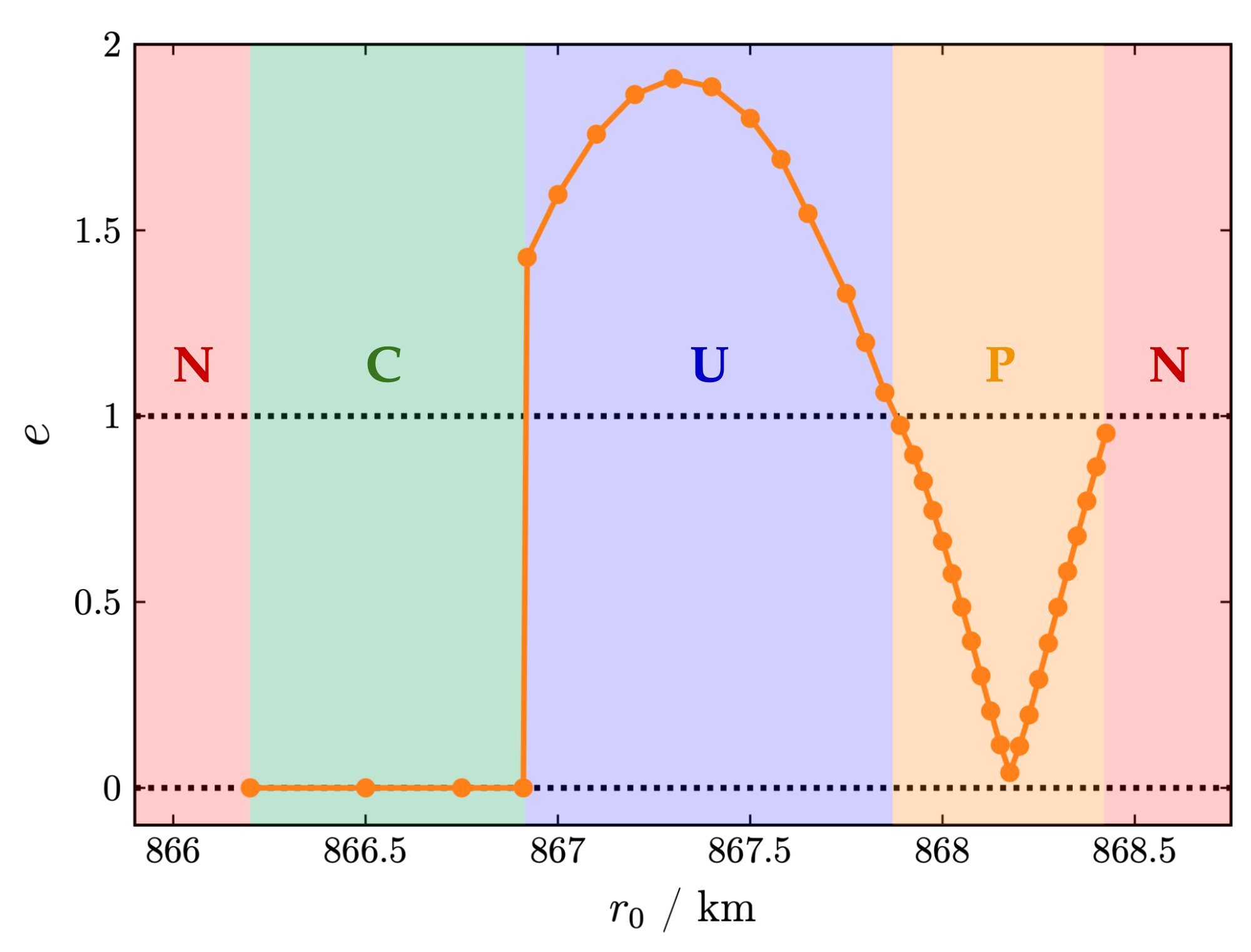}
    \caption{\footnotesize Eccentricity of resulting orbits with different turning radius $r_0$ in the interval shown in Fig. \ref{fig:anilloscond} for $f(R)=-0.05R^2$, $P_c=7\cdot10^{-4}$ km$^{-2}$ and the EoS \textit{Soft}. As expected, $e>1$ inside the \textit{unbounded} (\textbf{U}) region, $e<1$ inside the \textit{precessing} (\textbf{P}) region and $e=0$ inside the \textit{circular} (\textbf{C}) region. The \textit{no orbit} (\textbf{N}) regions have, of course, no eccentricity to be shown.}
    \label{fig:eccentricity}
\end{center}
\end{figure}

Note that each condition's impact on the orbit depends upon the previous ones, so stability $(C_3>0)$ means nothing if the orbit is unbounded $(C_2<0)$, and boundedness $(C_2>0)$ lacks significance if the orbit does not even exist in the first place $(C_1<0)$. A labelled illustration of these conditions and their consequences on the orbits is shown in Fig. \ref{fig:anilloscond}. Then, for different turning radii $r_0$ inside the labelled rings in that figure, some orbits have been computed and are shown in Fig. \ref{fig:orbitas}. Their eccentricities were calculated numerically and is shown in Fig. \ref{fig:eccentricity}.

\subsection{Geodesic equations in static and spherically symmetric spacetimes}
\label{SecIIIE}
As shown in the previous sections, not all radii are allowed for stable circular orbits in the $f(R)=aR^2$ model. Computing and plotting the resulting orbits inside and outside the \textit{SCO}s existence rings proves then to be of special interest, both to check that the previous analysis is correct and to explore the behaviour of the particles outside the existence rings. To carry out this analysis it will be necessary to solve the geodesic equations, which describe how massive or non-massive (light) particles move under the influence of the curvature of spacetime.

Taking the derivative of Eq. (\ref{eq:epsilon^2_f(R)}) with respect to $\phi$ and simplifying, we arrive at
\begin{equation}
    0=-k^2A^{-2}\frac{{\rm d}A}{{\rm d}\xi}-h^2\left[\frac{{\rm d}B}{{\rm d}\xi}\left(\frac{{\rm d}\xi}{{\rm d}\phi}\right)^2+2B\frac{{\rm d}^2\xi}{{\rm d}\phi^2}\right]-2h^2\xi,
\end{equation}
where we can substitute $\left(\frac{{\rm d}\xi}{{\rm d}\phi}\right)^2$ by isolating it from Eq. \eqref{eq:epsilon^2_f(R)}. Thus, the general geodesic equations become
\begin{equation}
    \frac{k^2}{h^2}\frac{\frac{{\rm d} A}{{\rm d} \phi}}{A^2}+\frac{\frac{d B}{d \phi}}{B}\left[\frac{k^2-A\epsilon^2}{Ah^2}-\xi^2\right]+2B\frac{{\rm d}^2\xi}{{\rm d}\phi^2}+2\xi=0\,,
    \label{eq:geodes_f(R)}
\end{equation}
together with Eqs. (\ref{EL-eqns}).

In the starting position of the particle $(r,\phi)=(r_\text{par},0)$ we take the value $\xi(0)=\frac{1}{r_\text{par}}$, so that the initial condition on the derivative becomes
\begin{equation}
    \frac{\dd\xi}{\dd\phi}(0)=\qty[\frac{k^2-A(r_\text{par})\epsilon^2}{A(r_\text{par})B(r_\text{par})h^2}-\frac{\xi(0)^2}{B(r_\text{par})}]^{1/2},
\end{equation}

Integrating numerically Eq. (\ref{eq:geodes_f(R)}) with these initial conditions and for different values of $h$ and $k$, we obtain the orbits of photons ($\epsilon=0$) or massive particles ($\epsilon=1$) for the outer spacetime of the model. Specifically, when forcing $h$ and $k$ to take the values given by Eqs. \eqref{eq:h^2} and \eqref{eq:k^2}, we obtain different behaviors depending on the chosen turning radius $r_0$ as shown in Fig. \ref{fig:anilloscond}. Some of the resulting orbits, one for each possible region, are shown in Fig. \ref{fig:orbitas}.

\section{Photon orbits in static and spherically symmetric spacetimes in $f(R)$ theories}
\label{orbits_photons}

The analysis for photon orbits is equivalent to that of massive particles, setting $\epsilon=0$ in Eq. \eqref{eq:epsilon^2_f(R)}:
\begin{equation}   
\frac{1}{2}AB\dot{r}^2+ V_{\rm eff}=\frac{1}{2}k^2\,,
\label{eq:energy_photons}
\end{equation}  
where now the effective potential takes the form
\begin{equation}  
\label{V_eff}  
V^\gamma_\text{eff}=\frac{1}{2}A\frac{h^2}{r^2}.
\end{equation}  

Imposing the condition of circularity ($\frac{{\rm d}V_\text{eff}}{{\rm d}r}\vert_{r_0}=0$) yields an equation involving only functions of $r_0$, such that its solutions are the only allowed radii, also called \textit{photon spheres}, as follows:
\begin{equation}
    r_0A'(r_0)-2A(r_0)=0.
    \label{eq:photonshpere}
\end{equation}
Should one plug the GR expression for the exterior metric coefficient $A(r)=1-\frac{2M}{r}$ into this equation, the well-known value for the photon sphere radius is recovered:
\begin{equation}
    r_\gamma^\text{GR}=3M.
    \label{eq:3M}
\end{equation}

\begin{figure}[htbp!]
\begin{center}
    \centering
    \includegraphics[width=0.75\linewidth]{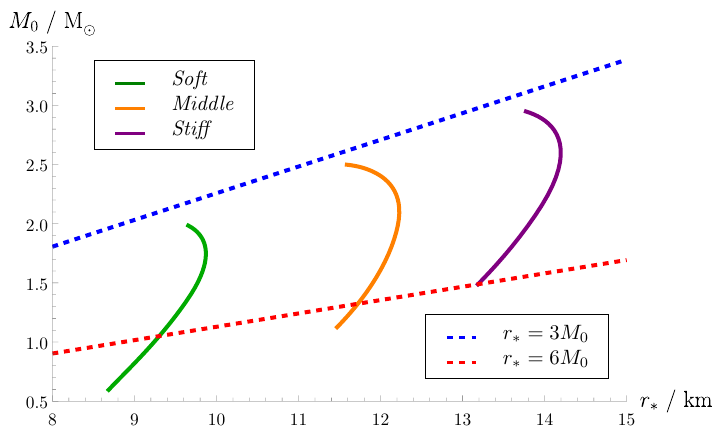}
    \caption{\footnotesize Mass-radius diagrams for all three studied EoS and a range of central pressures $P_C\in[5\cdot10^{-5},7\cdot10^{{-4}}]$ km. For the considered range of values of $|a|\in[1\cdot10^{-4},7,5\cdot10^{-3}]$ km, neither the bare mass $M_0$ nor the stellar radius $r_*$ change noticeably. The stars below $r_*=6M_0$ line have a low enough compactness for the ISCO to coincide with the stellar radius. Stars above the $r_*=3M_0$ line would be compact enough to allow for a photon sphere in GR.}
    \label{fig:MR_limits}
\end{center}
\end{figure}

Hence, for a photon sphere to exist in GR, the corresponding neutron star must have a radius smaller than this value, constituting a so-called {\it ultracompact object}, which is only realistically possible under specific conditions for some exceptional EoS \cite{Rezzolla_2026}. For the EoS and parameter-space region studied in this work, no such objects are found, as shown in Figure \ref{fig:MR_limits}.

\begin{figure}[H]
\begin{center}
    \centering
    \includegraphics[width=0.6\linewidth]{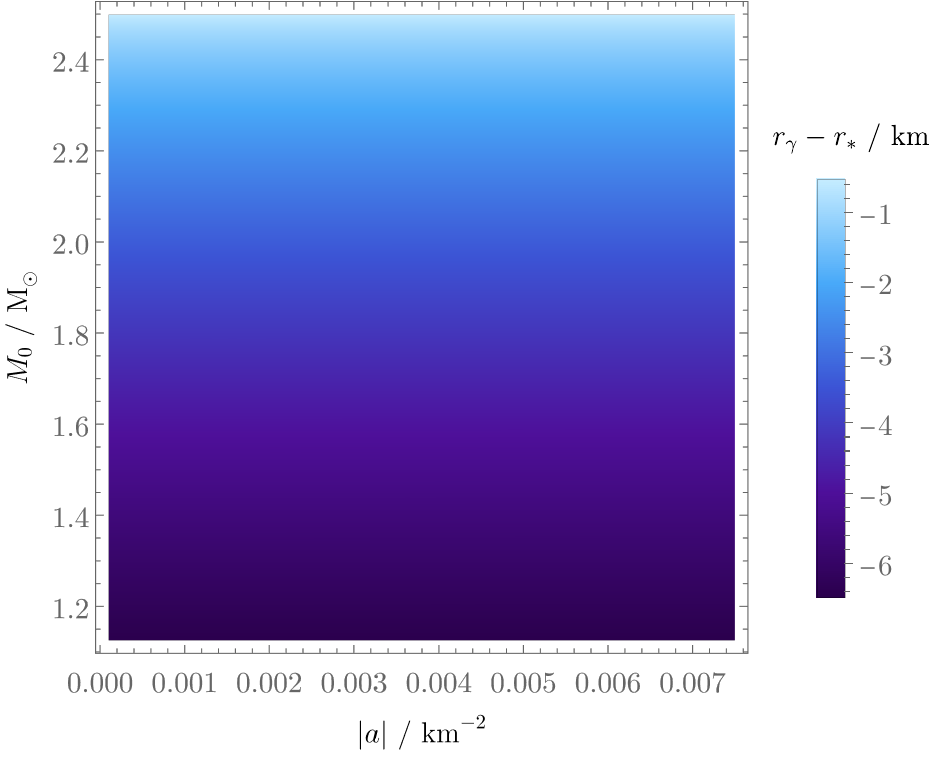}
    \caption{\footnotesize Difference between the photon sphere radius $r_\gamma$ and the stellar radius $r_*$ versus the bare stellar mass $M_0$ and parameter $|a|$ for the EoS \textit{Middle}. Since all values are negative in the studied region, no photon spheres are found outside the star.}
    \label{fig:nofotosfera}
\end{center}
\end{figure}

As we know, Eq. \eqref{eq:3M} does not hold for $f(R)$ theories. Nonetheless, for the EoS and the parameter-space regions studied in this work, we conclude the non existence of
photon spheres (solutions to Eq. \eqref{eq:photonshpere} outside the star). For illustrative purposes, we have exemplified this conclusion for the EoS \textit{Middle} in Fig. \ref{fig:nofotosfera}.

\section{Conclusions}
\label{Conclusions} 

In this work we investigated the structure and properties of massive and massless particle orbits in realistic static and spherically symmetric neutron-star spacetimes within metric $f(R)$ gravity, focusing on the quadratic Starobinsky model $f(R)=aR^2$ with $a<0$. The interior stellar configurations were obtained by numerically solving the modified Tolman--Oppenheimer--Volkoff system for several realistic neutron-matter equations of state and were consistently matched to the exterior vacuum solutions by enforcing the full set of junction conditions required in metric $f(R)$ theories, in particular the continuity of the metric, the extrinsic curvature, the Ricci scalar, and its normal derivative.

A central result of our analysis is that, unlike in General Relativity, the exterior spacetime is characterized by a nonvanishing, oscillatory Ricci scalar, which induces damped oscillations in the metric coefficients outside the stellar surface. These oscillations, originating from the massive scalar degree of freedom inherent to metric $f(R)$ gravity, have a direct and nontrivial impact on the geodesic structure of the spacetime.

We have shown that the existence and stability of circular geodesics for massive particles are in fact go\-ver\-ned by three radial conditions derived from the distinctive attributes of the effective potential. In contrast with the GR case---where all stable circular orbits exist beyond a single most stable circular orbit (ISCO) radius and this analysis is usually bypassed---the oscillatory nature of the exterior metric in $f(R)$ gravity generically leads to the appearance of discrete radial bands (dubbed {\it rings} herein) of stable circular orbits, separated by regions where no such orbits are allowed. Among these, a dominant \emph{principal ring} may arise, corresponding to the widest contiguous region of stability.
Also, we have assessed that the properties of this principal ring, including its existence and radial width, depend sensitively upon the neutron star’s central pressure, the equation of state, and the magnitude of the $f(R)$ parameter $|a|$. We find that increasing either the central pressure or the deviation from GR tends to suppress the principal ring, which may eventually disappear entirely. This demonstrates that even small departures from GR in the strong-field regime can produce qualitative changes in orbital stability.

Moreover, we have shown how the behavior of the innermost stable circular orbit (ISCO) is also significantly modified. Indeed, depending on the stellar configuration and the value of $a$, the ISCO in $f(R)$ gravity can lie either inside or outside its GR counterpart, and in some cases it may coincide with the stellar radius, allowing for stable circular motion arbitrarily close to the star surface. As a consequence, the ISCO ceases to be a universal quantity and becomes strongly theory - and configuration-dependent.

Beyond circular motion, we have explored the dynamics of massive particles in regions where one or more of the stability conditions fail. This analysis reveals a rich phenomenology, including bound but unstable precessing orbits, unbound trajectories, and forbidden regions with no turning-point solutions, fully consistent with the structure of the derived conditions.

Finally, we have analyzed null geodesics and shown that, for all equations of state and parameter ranges considered, no photon spheres exist outside the neutron star. Thus, despite the presence of scalar-induced oscillations in the exterior geometry, realistic neutron stars in the quadratic $f(R)$ model do not exhibit ultracompact behavior or light-trapping surfaces.

Overall, our results demonstrate that stable circular orbits around neutron stars provide a highly sensitive probe of modified gravity effects in the strong-field regime. The emergence of SCO rings, the modification of the ISCO, and the altered orbital landscape reported here constitute potentially observable signatures that could be relevant for accretion disk dy\-na\-mics and future precision tests of gravity in the vicinity of compact objects.

\vspace{1cm}
{\bf Acknowledgments:} 
 This work is supported by the Spanish Grants PID2024-158938NB-I00, PID2023-149560NB-C21, and the Severo Ochoa Excellence Grant CEX2023-001292-S, funded by MICIU/AEI/10.13039/501100011033 (“ERDF A way of making Europe”, “PGC Generacion de Conocimiento”) and FEDER, UE. The authors also acknowledge financial support from the project i-COOPB23096 (funded by CSIC). AdlCD acknowledges support from NRF Grant CSUR23042798041 and Project SA097P24 funded by Junta de Castilla y Le\'on (Spain). This work is also supported by CosmoVerse CA21136 and CaLISTA CA21109 COST actions, European Cooperation in Science and Technology.

\bibliography{bib-article}
\vspace{1cm}
\noindent\rule{0.48\textwidth}{0.5pt}

\appendix
\section{Curvature scalar checking and asymptotic mass}
\label{AppendinxA}

In the scalar-tensor picture we have the usual equi\-va\-lences $\phi\equiv 1+f_{R}$ for the scalar field and $V(\phi)\equiv R(\phi)\phi-f(R(\phi))$ for its potential \cite{SotiriouFaraoni2010}. From this, it is straightforward to conclude
\begin{equation}
    \frac{\dd V}{\dd\phi}=R\,,
\end{equation}
and the scalar field mass is given by
\begin{equation}
    m_\phi^2=\frac{1}{3}(\phi_cV_c^{\prime\prime}-V_c^{\prime}),
    \label{scalar_field_mass}
\end{equation}
where $\phi_c\equiv 1+ f_R(R_c)$ and $V_c\equiv V(\phi_c)$ are the \emph{asymptotic (cosmological) boundary values} of the scalar field and its potential, respectively. Here $R_c$ denotes the Ricci scalar of the background cosmology far from the local system, which fixes the boundary condition for $\phi$ in the weak-field analysis \cite{Olmo2007,Olmo2005}. In the linearized, approximately static regime, the local perturbation
$\varphi(\vec x)$ satisfies a screened Poisson equation whose solution can be written in terms of a Green function \cite{Olmo2005}:
\begin{equation}
    \varphi(\vec{x})=\frac{\kappa^2}{12\pi}\int \dd^3\vec{x}^{\prime}\frac{\rho(\vec{x}^{\prime})}{|\vec{x}-\vec{x}^{\prime}|}F(|\vec{x}-\vec{x}^{\prime}|).
    \label{eq:phi_appendix}
\end{equation}
Here we write the full scalar as $\phi(\vec x)=\phi_c+\varphi(\vec x)$, where
$\phi_c$ is the asymptotic (cosmological) value and $\varphi$ is the local deviation sourced by the star \cite{Olmo2005}. The energy density of the source over whose volume we integrate is denoted as $\rho$, and the function $F$ depends upon the scalar field mass in the following way:
\begin{equation}
\label{Expressions_of_F}
    F(|\vec{x}-\vec{x}^{\prime}|)=\begin{cases}{\rm e}^{-m_\varphi|\vec{x}-\vec{x}^{\prime}|}&\mathrm{if}\quad m_\varphi^2>0\\\\\cos(m_\varphi|\vec{x}-\vec{x}^{\prime}|)&\mathrm{if}\quad m_\varphi^2<0\end{cases}\quad.
\end{equation}
Thus, in order to get the expression of the scalar curvature $R$ to lowest order, we expand the scalar field around its cosmological value $\phi(\vec{x})=\phi_{c}+\varphi(\vec{x})$. Plugging in Eq. \ref{eq:phi_appendix}, we get
\begin{equation}
\begin{aligned}
    R&=V'(\phi_{c}+\varphi)\approx V'(\phi_{c})+V''(\phi_c)\,\varphi\\
    &=V'(\phi_{c})+V''(\phi_c)\,\frac{\kappa^2}{12\pi}\int \dd^3\vec{x}^{\prime}\frac{\rho(\vec{x}^{\prime})}{|\vec{x}-\vec{x}^{\prime}|}F(|\vec{x}-\vec{x}^{\prime}|)\,.
\end{aligned}
\label{eq:R_appendix}
\end{equation}
Now, considering the gravity Lagrangian $R+aR^2$, $\phi\equiv 1+2aR$ and hence $R=\frac{\phi-1}{2a}$, so that the potential and its derivatives read
\begin{equation}
    \begin{cases}
        V(\phi)=aR^2=\frac{(\phi-1)^2}{4a^2},\\
        V'(\phi)=R=\frac{\phi-1}{2a},\\
        V''(\phi)=\frac{1}{2a}.
    \end{cases}
\end{equation}
Plugging these into the expression for the scalar field mass \eqref{scalar_field_mass}, we see that for this $f(R)$ model it yields $m_\varphi^2= \frac{1}{6a}$, which is a constant.

Since in this work we have assumed $a<0$, it follows that $m_\varphi^2<0$ and, hence, $F(|\vec{x}-\vec{x}^{\prime}|)=\cos(m_\varphi|\vec{x}-\vec{x}^{\prime}|)$ as per Eq. \eqref{Expressions_of_F}. Substituting this into \ref{eq:R_appendix}, and assuming a Minkowskian background, so that $\phi_c\to1$,\footnote{Assuming an asymptotically flat background with $R_c=0$, the quadratic model $f(R)=R+aR^2$ implies $\phi_c=f_R(R_c)=1$ exactly, since $\phi=1+2aR$} we get that far from the star, the scalar curvature reads
\begin{equation}
\begin{aligned}
    R&\approx\cancel{\left.\frac{\phi_c - 1}{2a}\right|}_{\phi_c = 1}
    +\frac{\kappa^2}{24\pi a}\int \dd^3\vec{x}^{\prime}\frac{\rho(\vec{x}^{\prime})}{|\vec{x}-\vec{x}^{\prime}|}F(|\vec{x}-\vec{x}^{\prime}|)\\
    &=-\frac{\kappa^2}{24\pi|a|}\int \dd^3\vec{x}^{\prime}\frac{\rho(\vec{x}^{\prime})}{|\vec{x}-\vec{x}^{\prime}|}\cdot\cos\left(\frac{|\vec{x}-\vec{x}^{\prime}|}{\sqrt{6|a|}}\right).
\end{aligned}
\label{eq:R_appendix_bis}
\end{equation}
Given the spherical symmetry of the star, the integral above simplifies to a one-dimensional integral over the radial coordinate. Only the radius must therefore be integrated numerically, since the density profile $\rho(r)$ is obtained from the numerical solution described in the main body of this work. The resulting expression for the curvature scalar is
\begin{equation}
    R(r)\approx\frac{-\kappa^2}{\sqrt{24|a|}\,r}\int_0^{r_*}y\, \rho(y)\,\dd y\, \left[\sin{\frac{|r+y|}{\sqrt{6|a|}}-\sin{\frac{|r-y|}{\sqrt{6|a|}}}}\right].
\end{equation}
which can be simplified assuming $r>y$ and hence $|r-y|=r-y$, yielding
\begin{equation}
    R(r)\approx
-\frac{\kappa^{2}}{r\sqrt{6|a|}}
\cos\!\left(\frac{r}{\sqrt{6|a|}}\right)
\int_{0}^{r_*} y\,\rho(y)\,\!{\rm d}y\,
\sin\!\left(\frac{y}{\sqrt{6|a|}}\right)
\end{equation}
where the integral above reduces to a numerical value depending on the equation of state, the stellar radius and the value of $a$, so that $R(r)\propto \frac{\cos (|m_\varphi|r)}{r}$.

\section{Numerical consistency check in the scalar--tensor representation}\label{app:numcheck}

In the notation introduced in Appendix 
\ref{AppendinxA}, it's easy to check that the field equations take the form
\begin{equation}\label{eq:ST_field_eq}
\kappa\, T_{\mu\nu}=
\phi\, G_{\mu\nu}
+ \frac{1}{2} V(\phi)\, g_{\mu\nu}
- \nabla_\mu \nabla_\nu \phi
+ g_{\mu\nu}\, \Box \phi \, ,
\end{equation}
which is the scalar--tensor rewriting of Eq.~(\ref{fieldEQS}) in the main text. From here, by raising one index by contracting with the inverse metric,
\begin{equation}\label{eq:ST_field_eq_mixed}
\kappa\, T^{\mu}{}_{\nu}=
\phi\, G^{\mu}{}_{\nu}
+ \frac{1}{2} V(\phi)\, \delta^{\mu}{}_{\nu}
- \nabla^{\mu}\nabla_{\nu}\phi
+ \delta^{\mu}{}_{\nu}\,\Box\phi \, ,
\end{equation}
where $\delta^{\mu}{}_{\nu}=g^{\mu\alpha}g_{\alpha\nu}$. In the exterior region, $T^{\mu}{}_{\nu}=0$, and therefore right--hand side of Eq. (\ref{eq:ST_field_eq_mixed})
\begin{equation}
\mathcal{E}^{\mu}{}_{\nu}\equiv
\phi\, G^{\mu}{}_{\nu}
+ \frac{1}{2} V(\phi)\, \delta^{\mu}{}_{\nu}
- \nabla^{\mu}\nabla_{\nu}\phi
+ \delta^{\mu}{}_{\nu}\,\Box\phi
\end{equation}
must vanish identically.

As a direct numerical validation of the  integration in such exterior(matter vacuum) region, we evaluated $\mathcal{E}^{\mu}{}_{\nu}$ using the the numerical solutions
$A(r)$, $B(r)$ and $R(r)$ as determined in the bulk of the text and monitored the simplest independent component, namely $\mathcal{E}^{r}{}_{r}$. The explicit expression used in the check can be written as
\begin{widetext}
\begin{equation}
\mathcal{E}^r{}_r(r) =\;
\frac{2r\,A'(r)}{B(r)}
\Bigl(1+2aR(r)+a r\,R'(r)\Bigr)
- A(r)\Biggl[
2\Bigl(\frac{1}{B(r)}-1\Bigr)\Bigl(1+2aR(r)\Bigr)
+ a r\left(
r\,R(r)^2
+ \frac{8 R'(r)}{B(r)}\,
\right)
\Biggr]\,,
\end{equation}
\end{widetext}
where primes denote the usual derivatives with respect to the radial coordinate.

When represented along the full exterior integration domain, $\mathcal{E}^{r}{}_{r}(r)$ remains bounded and sufficiently compatible with zero. The residual oscillations are consistent with the expected accumulation of numerical discretization effects and interpolation errors in the reconstructed numerical functions and their derivatives. This provides an explicit check that the numerical exterior solutions satisfy the (vacuum) field equations to the required accuracy.

\end{document}